\documentclass[twocolumn]{aastex61}

\bibpunct{(}{)}{;}{a}{}{,} 
\usepackage{graphicx}
\usepackage{multirow}
\usepackage{float}
\usepackage{threeparttable}
\usepackage{natbib,twoopt}
\usepackage{url}
\usepackage{bm}
\usepackage{amssymb,amsmath}

\shorttitle{Clouds on 51~Eri~\lowercase{b}}
\shortauthors{Rajan et al.}

\received{March 26, 2017}
\revised{May 8, 2017}
\accepted{May 8, 2017}
\submitjournal{AJ}

\watermark{DRAFT}

\begin{document}

\title{Characterizing 51~Eri~\lowercase{b} from 1--5~$\mu$\lowercase{m}: a partly-cloudy exoplanet}

\correspondingauthor{Abhijith Rajan}
\email{arajan6@asu.edu}

\author[0000-0002-9246-5467]{Abhijith Rajan}
\affiliation{School of Earth and Space Exploration, Arizona State University, PO Box 871404, Tempe, AZ, USA 85287}

\author[0000-0003-0029-0258]{Julien Rameau}
\affiliation{Institut de Recherche sur les Exoplan{\`e}tes, D{\'e}partment de Physique, Universit{\'e} de Montr{\'e}al, Montr{\'e}al QC, Canada H3C 3J7}

\author[0000-0002-4918-0247]{Robert J. De Rosa}
\affiliation{Astronomy Department, University of California, Berkeley; Berkeley CA, USA 94720}

\author[0000-0002-5251-2943]{Mark S. Marley}
\affiliation{NASA Ames Research Center,  Mountain View, CA, USA 94035}

\author{James R. Graham}
\affiliation{Astronomy Department, University of California, Berkeley; Berkeley CA, USA 94720}

\author[0000-0003-1212-7538]{Bruce Macintosh}
\affiliation{Kavli Institute for Particle Astrophysics and Cosmology, Stanford University, Stanford, CA, USA 94305}

\author[0000-0002-4164-4182]{Christian Marois}
\affiliation{National Research Council of Canada Herzberg, 5071 West Saanich Rd, Victoria, BC, Canada V9E 2E7}
\affiliation{University of Victoria, 3800 Finnerty Rd, Victoria, BC, Canada V8P 5C2}

\author[0000-0002-4404-0456]{Caroline Morley}
\affiliation{Department of Astronomy, Harvard University, Cambridge, MA, USA 02138}

\author{Jennifer Patience}
\affiliation{School of Earth and Space Exploration, Arizona State University, PO Box 871404, Tempe, AZ, USA 85287}

\author{Laurent Pueyo}
\affiliation{Space Telescope Science Institute, Baltimore, MD, USA 21218}

\author[0000-0001-6800-3505]{Didier Saumon}
\affiliation{Los Alamos National Laboratory, P.O. Box 1663, MS F663, Los Alamos, NM USA, 87545}

\author[0000-0002-4479-8291]{Kimberly Ward-Duong}
\affiliation{School of Earth and Space Exploration, Arizona State University, PO Box 871404, Tempe, AZ, USA 85287}

\author[0000-0001-5172-7902]{S. Mark Ammons}
\affiliation{Lawrence Livermore National Laboratory, Livermore, CA, USA 94551}

\author[0000-0001-6364-2834]{Pauline Arriaga}
\affiliation{Department of Physics \& Astronomy, University of California, Los Angeles, CA, USA 90095}

\author[0000-0002-5407-2806]{Vanessa P. Bailey}
\affiliation{Kavli Institute for Particle Astrophysics and Cosmology, Stanford University, Stanford, CA, USA 94305}

\author[0000-0002-7129-3002]{Travis Barman}
\affiliation{Lunar and Planetary Laboratory, University of Arizona, Tucson AZ, USA 85721}

\author{Joanna Bulger}
\affiliation{Subaru Telescope, NAOJ, 650 North A{'o}hoku Place, Hilo, HI 96720, USA}

\author[0000-0002-3099-5024]{Adam S. Burrows}
\affiliation{Department of Astrophysical Sciences, Princeton University, Princeton, NJ, USA 08544}

\author[0000-0001-6305-7272]{Jeffrey Chilcote}
\affiliation{Dunlap Institute for Astronomy \& Astrophysics, University of Toronto, Toronto, ON, Canada M5S 3H4}

\author[0000-0003-0156-3019]{Tara Cotten}
\affiliation{Department of Physics and Astronomy, University of Georgia, Athens, GA, USA 30602}

\author[0000-0002-1483-8811]{Ian Czekala}
\affiliation{Kavli Institute for Particle Astrophysics and Cosmology, Stanford University, Stanford, CA, USA 94305}

\author{Rene Doyon}
\affiliation{Institut de Recherche sur les Exoplan{\`e}tes, D{\'e}partment de Physique, Universit{\'e} de Montr{\'e}al, Montr{\'e}al QC, Canada H3C 3J7}

\author[0000-0002-5092-6464]{Gaspard Duch{\^e}ne}
\affiliation{Astronomy Department, University of California, Berkeley; Berkeley CA, USA 94720}
\affiliation{Univ. Grenoble Alpes/CNRS, IPAG, F-38000 Grenoble, France}

\author[0000-0002-0792-3719]{Thomas M. Esposito}
\affiliation{Astronomy Department, University of California, Berkeley; Berkeley CA, USA 94720}

\author[0000-0002-0176-8973]{Michael P. Fitzgerald}
\affiliation{Department of Physics \& Astronomy, University of California, Los Angeles, CA, USA 90095}

\author[0000-0002-7821-0695]{Katherine B. Follette}
\affiliation{Kavli Institute for Particle Astrophysics and Cosmology, Stanford University, Stanford, CA, USA 94305}

\author[0000-0002-9843-4354]{Jonathan J. Fortney}
\affiliation{Department of Astronomy, UC Santa Cruz, 1156 High Street, Santa Cruz, CA, USA 95064}

\author[0000-0002-4144-5116]{Stephen J. Goodsell}
\affiliation{Gemini Observatory, 670 N. A'ohoku Place, Hilo, HI, USA 96720}

\author[0000-0002-7162-8036]{Alexandra Z. Greenbaum}
\affiliation{Department of Astronomy, University of Michigan, Ann Arbor MI, USA 48109}

\author[0000-0003-3726-5494]{Pascale Hibon}
\affiliation{Gemini Observatory, Casilla 603, La Serena, Chile}

\author[0000-0003-1498-6088]{Li-Wei Hung}
\affiliation{Department of Physics \& Astronomy, University of California, Los Angeles, CA, USA 90095}

\author{Patrick Ingraham}
\affiliation{Large Synoptic Survey Telescope, 950N Cherry Av, Tucson, AZ, USA 85719}

\author[0000-0003-4851-7706]{Mara Johnson-Groh}
\affiliation{University of Victoria, 3800 Finnerty Rd, Victoria, BC, Canada V8P 5C2}

\author{Paul Kalas}
\affiliation{Astronomy Department, University of California, Berkeley; Berkeley CA, USA 94720}
\affiliation{SETI Institute, Carl Sagan Center, 189 Bernardo Avenue,  Mountain View, CA, USA 94043}

\author[0000-0002-9936-6285]{Quinn Konopacky}
\affiliation{Center for Astrophysics and Space Science, University of California San Diego, La Jolla, CA, USA 92093}

\author{David Lafreni{\`e}re}
\affiliation{Institut de Recherche sur les Exoplan{\`e}tes, D{\'e}partment de Physique, Universit{\'e} de Montr{\'e}al, Montr{\'e}al QC, Canada H3C 3J7}

\author{James E. Larkin}
\affiliation{Department of Physics \& Astronomy, University of California, Los Angeles, CA, USA 90095}

\author{J{\'e}r{\^o}me Maire}
\affiliation{Center for Astrophysics and Space Science, University of California San Diego, La Jolla, CA, USA 92093}

\author[0000-0001-7016-7277]{Franck Marchis}
\affiliation{SETI Institute, Carl Sagan Center, 189 Bernardo Avenue,  Mountain View, CA, USA 94043}

\author[0000-0003-3050-8203]{Stanimir Metchev}
\affiliation{Department of Physics and Astronomy, Centre for Planetary Science and Exploration, The University of Western Ontario, London, ON N6A 3K7, Canada}
\affiliation{Department of Physics and Astronomy, Stony Brook University, Stony Brook, NY 11794-3800, USA}

\author[0000-0001-6205-9233]{Maxwell A. Millar-Blanchaer}
\affiliation{Jet Propulsion Laboratory, California Institute of Technology, Pasadena, CA, USA 91125}
\affiliation{NASA Hubble Fellow}

\author[0000-0002-1384-0063]{Katie M. Morzinski}
\affiliation{Steward Observatory, University of Arizona, Tucson AZ, USA 85721}

\author[0000-0001-6975-9056]{Eric L. Nielsen}
\affiliation{SETI Institute, Carl Sagan Center, 189 Bernardo Avenue,  Mountain View, CA, USA 94043}
\affiliation{Kavli Institute for Particle Astrophysics and Cosmology, Stanford University, Stanford, CA, USA 94305}

\author[0000-0001-7130-7681]{Rebecca Oppenheimer}
\affiliation{Department of Astrophysics, American Museum of Natural History, New York, NY, USA 10024}

\author{David Palmer}
\affiliation{Lawrence Livermore National Laboratory, Livermore, CA, USA 94551}

\author[0000-0002-5025-6827]{Rahul I. Patel}
\affiliation{Infrared Processing and Analysis Center, California Institute of Technology, Pasadena, CA, USA 91125}

\author[0000-0002-3191-8151]{Marshall Perrin}
\affiliation{Space Telescope Science Institute, Baltimore, MD, USA 21218}

\author{Lisa Poyneer}
\affiliation{Lawrence Livermore National Laboratory, Livermore, CA, USA 94551}

\author[0000-0002-9667-2244]{Fredrik T. Rantakyr{\"o}}
\affiliation{Gemini Observatory, Casilla 603, La Serena, Chile}

\author[0000-0003-2233-4821]{Jean-Baptiste Ruffio}
\affiliation{Kavli Institute for Particle Astrophysics and Cosmology, Stanford University, Stanford, CA, USA 94305}

\author[0000-0002-8711-7206]{Dmitry Savransky}
\affiliation{Sibley School of Mechanical and Aerospace Engineering, Cornell University, Ithaca, NY, USA 14853}

\author{Adam C. Schneider}
\affiliation{School of Earth and Space Exploration, Arizona State University, PO Box 871404, Tempe, AZ, USA 85287}

\author[0000-0003-1251-4124]{Anand Sivaramakrishnan}
\affiliation{Space Telescope Science Institute, Baltimore, MD, USA 21218}

\author[0000-0002-5815-7372]{Inseok Song}
\affiliation{Department of Physics and Astronomy, University of Georgia, Athens, GA, USA 30602}

\author[0000-0003-2753-2819]{R{\'e}mi Soummer}
\affiliation{Space Telescope Science Institute, Baltimore, MD, USA 21218}

\author{Sandrine Thomas}
\affiliation{Large Synoptic Survey Telescope, 950N Cherry Av, Tucson, AZ, USA 85719}

\author{Gautam Vasisht}
\affiliation{Jet Propulsion Laboratory, California Institute of Technology, Pasadena, CA, USA 91125}

\author{J. Kent Wallace}
\affiliation{Jet Propulsion Laboratory, California Institute of Technology, Pasadena, CA, USA 91125}

\author[0000-0002-4918-0247]{Jason J. Wang}
\affiliation{Astronomy Department, University of California, Berkeley; Berkeley CA, USA 94720}

\author{Sloane Wiktorowicz}
\affiliation{The Aerospace Corporation, El Segundo, CA, USA 90245}

\author[0000-0002-9977-8255]{Schuyler Wolff}
\affiliation{Department of Physics and Astronomy, Johns Hopkins University, Baltimore, MD, USA 21218}

\begin{abstract}

We present spectro-photometry spanning 1--5 $\mu$m of 51 Eridani b, a 2--10~M$_\text{Jup}$ planet discovered by the Gemini Planet Imager Exoplanet Survey. In this study, we present new $K1$~(1.90--2.19~$\mu$m) and $K2$~(2.10--2.40~$\mu$m) spectra taken with the Gemini Planet Imager as well as an updated $L_P$~(3.76~$\mu$m) and new $M_S$~(4.67~$\mu$m) photometry from the NIRC2 Narrow camera. The new data were combined with $J$~(1.13--1.35~$\mu$m) and $H$~(1.50--1.80~$\mu$m) spectra from the discovery epoch with the goal of better characterizing the planet properties. 51~Eri~b photometry is redder than field brown dwarfs as well as known young T-dwarfs with similar spectral type (between T4--T8) and we propose that 51 Eri b might be in the process of undergoing the transition from L-type to T-type. We used two complementary atmosphere model grids including either deep iron/silicate clouds or sulfide/salt clouds in the photosphere, spanning a range of cloud properties, including fully cloudy, cloud free and patchy/intermediate opacity clouds. Model fits suggest that 51~Eri~b has an effective temperature ranging between 605--737 K, a solar metallicity, a surface gravity of $\log$(g) = 3.5--4.0 dex, and the atmosphere requires a patchy cloud atmosphere to model the SED. From the model atmospheres, we infer a luminosity for the planet of -5.83 to -5.93 ($\log L/L_{\odot}$), leaving 51~Eri~b in the unique position as being one of the only directly imaged planet consistent with having formed via cold-start scenario. Comparisons of the planet SED against warm-start models indicates that the planet luminosity is best reproduced by a planet formed via core accretion with a core mass between 15 and 127 M$_{\oplus}$. 
\end{abstract}


\section{Introduction}

Until recently, most of the imaged planetary mass companions detected were typically orbiting their parent star at large orbital separations, $>$30~au. However, new instrumentation with second generation adaptive optics such as the Gemini Planet Imager \citep[GPI,][]{Macintosh:2014} and Spectro-Polarimetric High-contrast Exoplanet REsearch \citep[SPHERE,][]{Beuzit:2008} are now routinely obtaining deep contrasts ($>10^5-10^6$) in the inner arcsecond (5--30~au). The recent detection of new companions \citep{Macintosh:2015, Konopacky:2016, Wagner:2016, Milli:2017} and debris disks \citep{Currie:2015,Wahhaj:2016,Blanchaer:2016,Bonnefoy:2017} showcase the advances made by these next generation AO systems. Direct imaging, unlike non-direct methods such as radial velocity and transits, measures light from companions directly, which permits measuring the atmospheric spectrum, with the caveat that the final calibration is dependant on complete understanding of the stellar properties. These new AO instruments combine excellent image stability and high throughput with IFU spectrographs, enabling the measurement of a spectrum of the planet in the near infrared (IR) wavelength range. Combining the near-IR spectra with mid-IR photometry from instruments such as Keck/NIRC2, MagAO/Clio or LBT/LMIRCam, provides valuable constraints on the effective temperature and non-equilibrium chemistry when undertaking comprehensive modeling of the exoplanet spectral energy distribution.

In this study we focus on the planetary companions, 51~Eridani~b \citep[51~Eri~b;][]{Macintosh:2015}. 51~Eri~b is the first planet discovered by the Gemini Planet Imager Exoplanet Survey (GPIES), a survey targeting 600 young and nearby stars using GPI to search for exoplanets. The planet orbits 51~Eri~A, a young F0IV star that is part of the $\beta$~Pic moving group \citep{Zuckerman:2001}. In this study, we adopt an age of of $26\pm3$~Myr for the $\beta$~Pic moving group \citep{Nielsen:2016}. However, the age of the group is a topic of considerable debate and has been revised several times e.g. $21\pm4$ \citep{Binks:2014}, $23\pm3$ \citep{Mamajek:2014}, $20\pm6$ \citep{Macintosh:2015}, $24\pm3$ \citep{Bell:2015}. The primary is part of a hierarchical triple with two M-star companions, GJ~3305AB, separated from the primary by $\sim$2000~au \citep{Feigelson:2006, Kasper:2007, Montet:2015}. 51~Eri~A is known to have an IR excess, and a debris disk was detected in \emph{Herschel Space Observatory} 70 and 100 $\mu$m bands with very low IR luminosity of L$_\text{IR}$/L$_{\star} = 2 \times 10^{-6}$ and an a lower limit on the inner radius of 82~au \citep{Marichalar:2014} as well as a detection at 24 $\mu$m with the \emph{Spitzer Space Telescope} \citep{Rebull:2008}. The debris disk was not detected in \citet{Macintosh:2015}, which, given the low fractional luminosity would be extremely challenging. The analysis of the atmosphere of 51~Eri~b by \citet{Macintosh:2015} was based on GPI $JH$ spectra (1.1--1.8~$\mu$m) and Keck $L_P$ photometry (3.76~$\mu$m), using two different model atmosphere grids to estimate planet properties. While the models agreed on the temperature and luminosity, they were highly discrepant in terms of best fitting surface gravity with one grid suggesting low surface gravity and youth while the other required a high surface gravity and an old planet. Similarly, one grid best fit the atmosphere when using a linear combination of cloudy and clear models while the other best fit the data with clear atmosphere. These discrepancies indicate that more data is required to fully constrain the planet parameters. 

In this paper, we present new observations and revised data analysis that can be used to discriminate between some of the disagreements. In Section~\ref{sec:obs}, we present the first $K1$ (1.90--2.19~$\mu$m) and $K2$ (2.10--2.40~$\mu$m) spectrum of the planet taken with GPI. We also present updated $L_P$ photometry and new observations of the planet in the $M_S$-band (4.67~$\mu$m). In Section~\ref{sec:res}, we present new near-IR photometry of the star and revise the stellar spectral energy distribution (SED) used in the rest of the analysis. In Section~\ref{sec:ana}, we examine the near- and mid-IR photometry of 51~Eri~b in relation to that of other field and young brown dwarfs through the brown dwarf color-magnitude diagram. We also compare the near-IR spectrum of 51~Eri~b to field brown dwarfs, and planetary-mass companions to estimate the best fitting spectral type of the planet. Finally, in Section~\ref{model} we model the planet SED using two different grids spanning effective temperatures from 450K to 1000K with deep iron/silicate clouds or sulfide/salt clouds. The 1--5~$\mu$m spectral energy distribution in combination with these two model grids with help refine the planet properties and clarify whether the atmosphere is best fit by clouds, or not. 

\begin{deluxetable*}{cccccccc}
\tablecolumns{8} 
\tablewidth{0pt} 
\tablecaption{ Observations of 51~Eri~b \label{tab:phot}} 
\tablehead{ 
\colhead{Date}  & \colhead{Instrument} & \colhead{Filter} & \colhead{Total Int.} & \colhead{Field} & \colhead{Averaged} & \colhead{Averaged} & \colhead{Averaged} \\ 
                &      &            &  \colhead{time (min)} & \colhead{Rot. (deg)} & \colhead{airmass} & \colhead{DIMM seeing} (as) & \colhead{MASS $\tau_0$ (ms)}}
\startdata 
2015 Jan 30  & GS/GPI & $J$\tablenotemark{a}  & 70 & 23.8 & 1.15 & 0.52 & 3.26 \\
2014 Dec 18  & GS/GPI & $H$\tablenotemark{a}  & 38 & 37.7 & 1.14 & --   & --   \\
2015 Nov 06  & GS/GPI & $K1$\tablenotemark{b} & 55 & 30.5 & 1.17 & 0.38 & 1.56 \\
2015 Dec 18  & GS/GPI & $K2$\tablenotemark{b} & 103 & 71.7 & 1.22 & 0.69 & 0.94 \\
2016 Jan 28  & GS/GPI & $K1$\tablenotemark{b} & 97 & 55.5 & 1.15 & 0.86 & 4.40 \\
2015 Oct 27  & Keck/NIRC2 & $L_P$\tablenotemark{b} & 100 & 74.2 & 1.10 & -- & -- \\
2016 Jan 02  & Keck/NIRC2 & $M_S$\tablenotemark{b} & 139 & 115.7 & 1.18 & -- & -- \\
2016 Jan 21  & Keck/NIRC2 & $M_S$\tablenotemark{b} & 174 & 116.0 & 1.21 & -- & -- \\
2016 Feb 04  & Keck/NIRC2 & $M_S$\tablenotemark{b} & 148 & 101.4 & 1.21 & -- & -- \\
2016 Feb 05  & Keck/NIRC2 & $M_S$\tablenotemark{b} & 142 & 102.1 & 1.21 & -- & -- \\
\enddata 
\tablenotetext{a}{\citet{Macintosh:2015}}
\tablenotetext{b}{This work}
\end{deluxetable*}

\section{Observations and Data Reduction}
\label{sec:obs}

\subsection{GPI K1 and K2}
51~Eri~b was observed with the Integral Field Spectrograph (IFS) of GPI through the $K1$ filter on 2015 November 06 UT and 2016 January 28 and through the $K2$ filter on 2015 December 18 UT (see Table~\ref{tab:phot}). Standard procedures, namely using an argon-arc lamp, were used to correct the data for instrumental flexure. To maximize the parallactic rotation for Angular Differential Imaging \citep[ADI;][]{Marois:2006a}, the observations were centered on meridian passage. All the GPI datasets underwent the same initial data processing steps using the GPI Data Reduction Pipeline v1.3.0 \citep[DRP;][]{Perrin:2014}. The processing steps included dark current subtraction, bad pixel identification and interpolation, this is followed by compensating for instrument flexure using the argon arc spectrum \citep{Wolff:2014}. Following this step, the microspectra are extracted to generate the IFS datacubes \citep{Maire:2014}. During the process of generating the 3D (x, y, $\lambda$) cubes, the microspectra data are resampled to $\lambda/\delta \lambda$ = 65, and 75 at $K1$ and $K2$, respectively, after which they are interpolated to a common wavelength scale and corrected for geometric distortion \citep{Konopacky:2014}. The datacubes are then aligned to a common center calculated using the four satellite spots \citep{Wang:2014}. The satellite spots are copies of the occulted central star, generated by the use of a regular square grid printed on the apodizer in the pupil plane \citep{Sivaramakrishnan:2006, Marois:2006b, Macintosh:2014}. The satellite spots also help convert the photometry from contrast units to flux units. No background subtraction was performed since the following steps of high-pass filtering and PSF subtraction efficiently remove this low frequency component.

Further steps to remove quasi-static speckles and large scale structures were executed outside the DRP. Each datacube was filtered using an unsharp mask with a box width of 11 pixels. The four satellite spots were then extracted from each wavelength slice, and averaged over time to obtain templates of star point spread function (PSF). The Linear Optimized Combination of Images algorithm \citep[LOCI,][]{Lafreniere:2007} was used to suppress the speckle field in each frame using a combination of aggressive parameters: $dr=5$ px, $N_A$=200 PSF full width at half maximum (FWHM), $g=0.5$, and $N_\delta=0.5-0.75$ FWHM for the three datasets. Where $dr$ is the radial width of the optimization zone, $N_A$ is the number of PSF FWHM that can be included in the zone, $g$ is the ratio of the azimuthal and radial widths of the optimization zone, and $N_\delta$ defines the maximum separation of a potential astrophysical source in FWHM between the target and the reference PSF. The residual image of each wavelength slice was built from a trimmed ($10\%$) temporal average of the sequence.

\begin{figure}
\centering
\includegraphics[width=4.0cm]{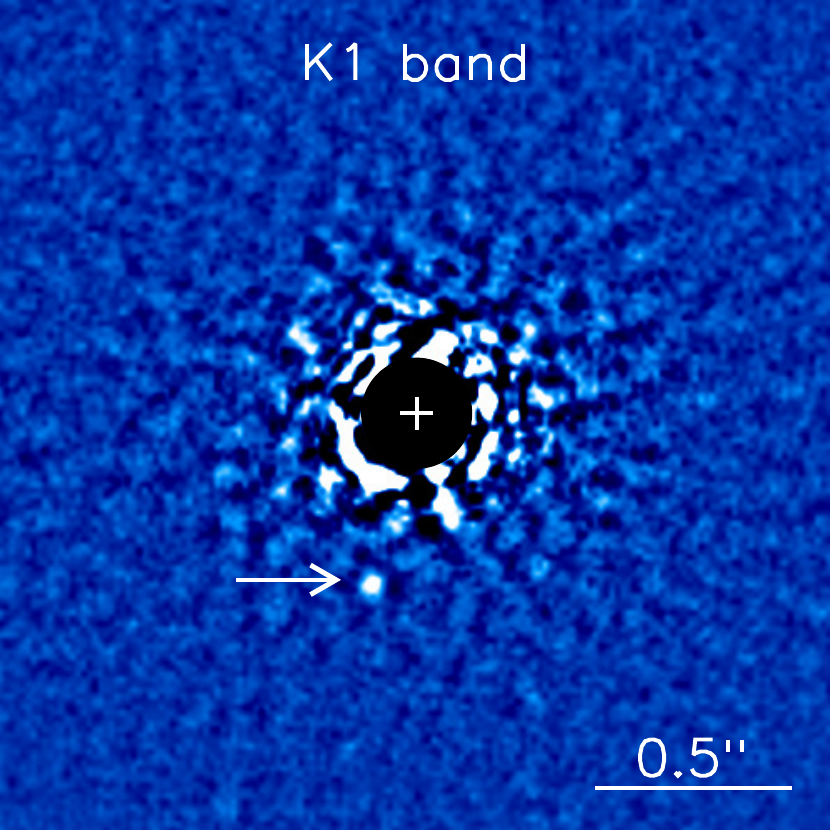}\includegraphics[width=4.0cm]{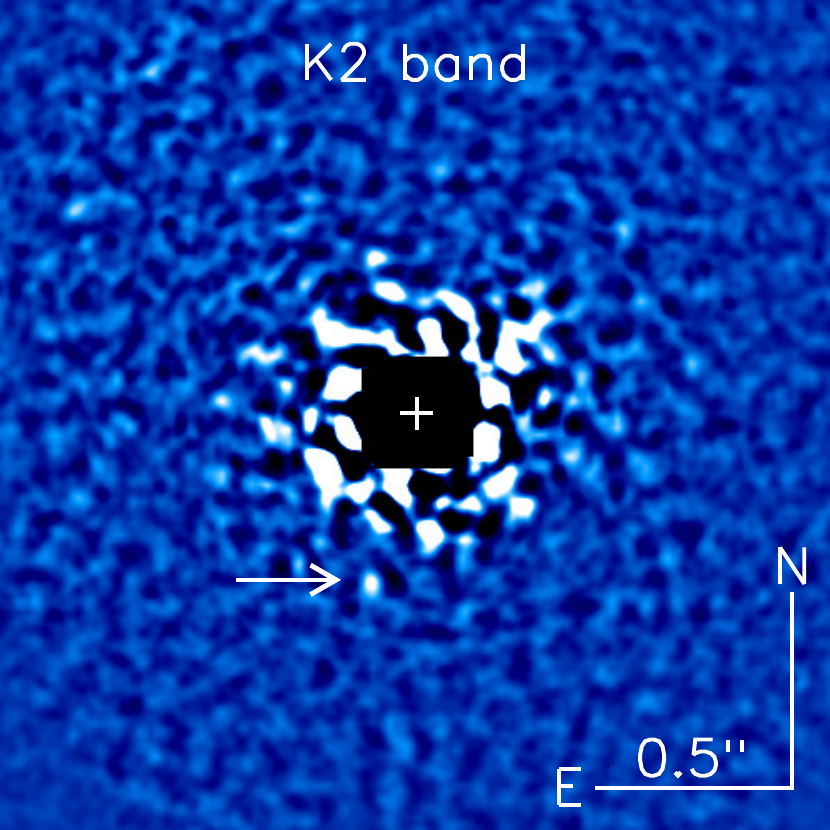}\\
\includegraphics[width=4.0cm]{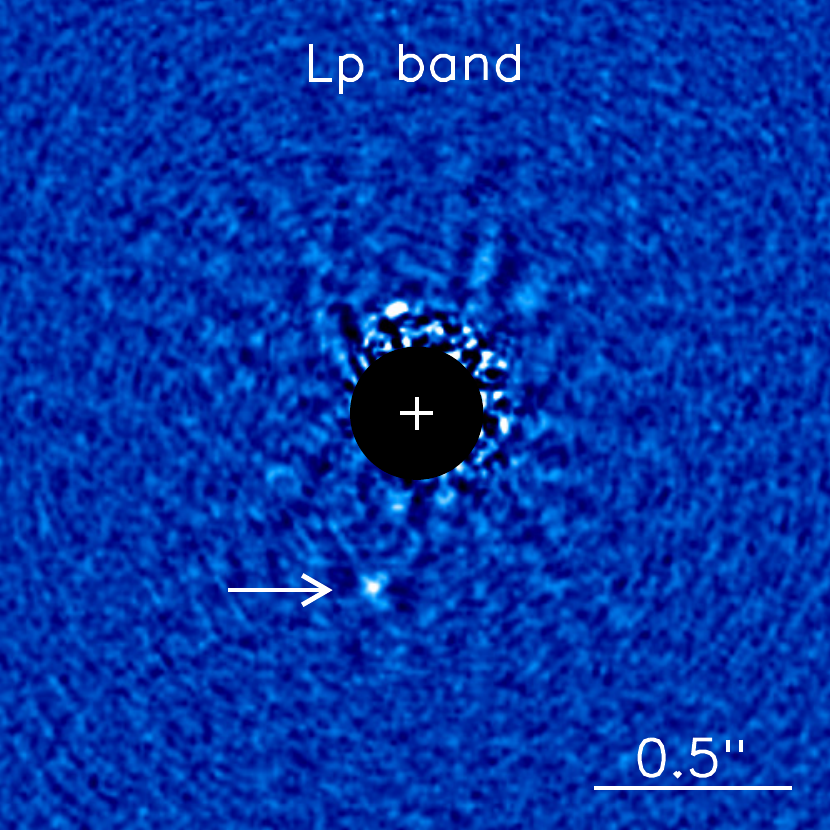}\includegraphics[width=4.0cm]{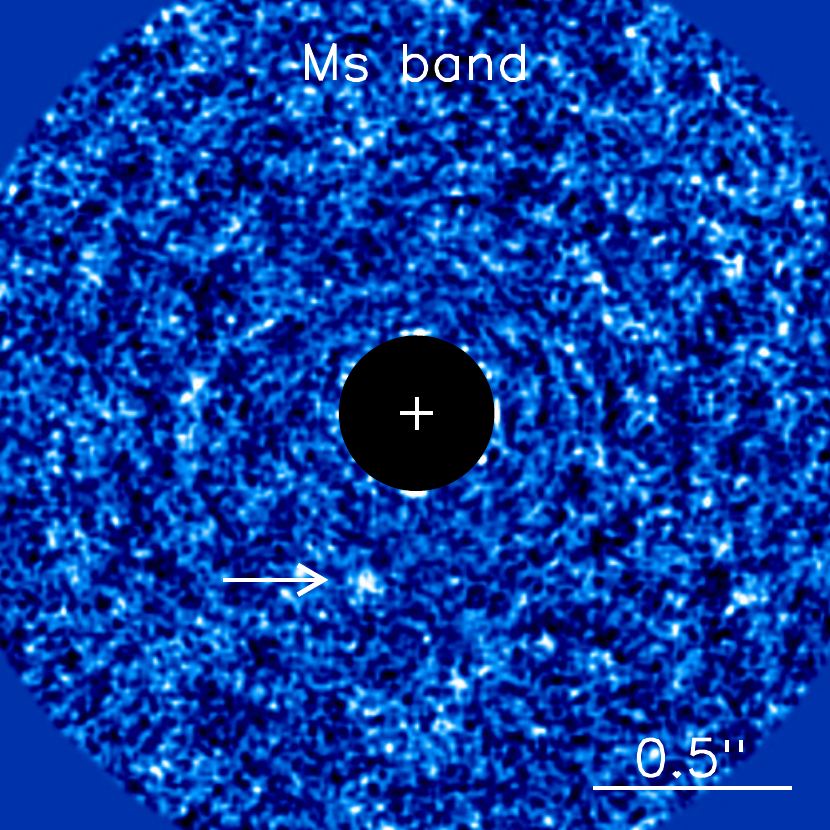}
\caption{Final PSF subtracted images of 51~Eri~b. (Top) LOCI-reduced GPIES images at $K1$ (2016 Jan 28, left) and $K2$ band (2015 Dec 18, right). (Bottom) {\tt pyKLIP}-reduced NIRC2 images, smoothed with a box of width of 2 pixels, at $L_P$ (2015 Oct 27, left), and a combined image of all four $M_S$ datasets (right). The images are scaled linearly, but are different in each panel in order to saturate the core of the planet PSF. \label{fig:gpiesK1K2LpMs}}
\end{figure}
 
Final $K1$ and $K2$ broad-band images were created using a weighted-mean of the residual wavelength frames according to the spectrum of the planet, examples of which can be found in Figure \ref{fig:gpiesK1K2LpMs}. These broad-band images were used to extract the astrometry of the planet in each dataset thanks to higher signal-to-noise ratio (SNR) than in individual frames. To do so, a negative template PSF was injected into the raw data at the estimated position and flux of the planet before applying LOCI and reduced using the same matrix coefficients as the original reduction \citep{Marois:2010}. The process was iterated over these three parameters (x position, y position, flux) with the amoeba-simplex optimization \citep{NelderMead:1965} until the integration squared pixel noise in a wedge of 2$\times$2 FWHM was minimized. The best fit position was then used to extract the contrast of the planet in each dataset. The same procedure was executed in the non-collapsed wavelength residual images but varying only the flux of the negative template PSF and keeping the position fixed to prevent the algorithm from catching nearby brighter residual speckles in the lower SNR spectral slices. To measure uncertainties, we injected the template PSF with the measured planet contrast into each datacube at the same separation and 20 different position angles. We measured the fake signal with the same extraction procedure. The contrasts measured in the 2015 Nov 06 and 2016 Jan 28 $K1$ datasets agreed within the uncertainties, the latter having significantly better SNR, and were combined with weighted mean to provide the final planet contrasts.
 
\subsubsection{Spectral covariances}
\label{sec:covariance}
Estimation of a directly imaged planets properties from its measured spectrum is complicated by the fact that spectral covariances are present within the extracted spectra. In the GPI data these are caused by the residual speckle noise in the final PSF-subtracted image, and the oversampling of the individual microspectra during the initial data reduction process. Atmosphere modeling without properly accounting for these covariances can lead to biased results. We present the derivation of the correlation using the parameterization of \citet{Greco:2016} in the Appendix~\ref{app:cov2}.

We use the spectral covariance when carrying out comparison of the planet spectrophotometry against other field and young dwarfs as well as during model fitting. The covariance helps correctly account for the correlation in the spectra while also increasing the importance of the photometry, and thus the use of the covariance tends to move the best fits towards cooler temperatures when compared to using the variance directly. 

\subsection{Keck $L_P$}
We observed the 51~Eri system on 2015 Oct 27 in the $L_P$ filter with the NIRC2 camera \citep{McLean:2003} at the Keck-II observatory (Program ID - U055N2). The observations were taken in ADI mode, starting $\sim$1~hour prior to meridian crossing to maximize the field of view rotation. The target was observed for $\sim$3~hours total, with 100~min of on-source integration. The observations were acquired using the 400mas focal plane mask and the circular undersized ``incircle'' cold stop. To calibrate the planet brightness unsaturated observations of the star were taken at the end of the observing sequence. The images were dark and flat field corrected. We used the $K_S$-band lamp flats to build the flatfield and masked hot and bad pixels. As these observations were taken after the April 2015 servicing of NIRC2, the geometric distortion was corrected using the solution presented in \citet{Service:2016} (updating the original \citet{Yelda:2010} solution), with an updated plate scale of 9.971$\pm$0.004 mas pixel$^{-1}$ and the offset angle $\beta$ ($0.262\pm0.020$) which is required when calculating the position angle prior to rotating the images to put north up \citep{Yelda:2010}. Post-processing of the data was carried out using the Python version of the Karhunen-Lo\`eve Image Projection algorithm \citep[KLIP,][]{Soummer:2012, Amara:2012}, \texttt{pyKLIP} \citep{Wang:2015}. As part of this study, we included a NIRC2 module in the \texttt{pyKLIP} codebase that is publicly available for users. \footnote{\url{https://bitbucket.org/pyKLIP/pyklip}} The algorithm accepts aligned images and performs PSF subtraction using KLIP where the image can be divided into sections both radially and azimuthally. Aside from the choice of zones, there are two main parameters that were adjusted, the number of modes used in the Karhunen?-Lo\`eve (KL) transform and an exclusion criterion for reference PSFs, similar to $N_\delta$ mentioned above, that determines the number of pixels an astrophysical source would move due to the rotation of the reference stack. We carried out a parameter search where the four parameters mentioned were varied to optimize the signal to noise in the planet signal. The planet photometry was estimated using the method described above for the $K1$ and $K2$ filters, using a negative template PSF. The $L_P$ magnitude contrast for the star-planet is 11.58$\pm$0.15 mag which agrees very well with the photometry in the original epoch, 11.62$\pm$0.17 mag. The weighted mean of both measurements is used in the rest of the analysis.

\subsection{Keck $M_S$}
Observations of 51~Eri~b were taken in the $M_S$-band filter over four separate half nights on 2016 Jan 02, 21 and 2016 Feb 04, 05 with Keck/NIRC2 Narrow camera. The details of the observations are presented in Table~\ref{tab:phot}. Each night the target was observed for a period of $\sim$6 hours, as part of two separate NASA and UC Keck observing programs (Program ID - N179N2, U117N2). The data were obtained in ADI mode, with the field of view rotating at the sidereal rate. To reduce the effects of persistence and enable accurate thermal background correction, the star was nodded across the detector in four large dithers centered in each quadrant of the detector. Furthermore to prevent saturation of the detector by the thermal background, the exposures were limited to 0.3s with 200 co-adds, without using an occulting spot. The images were dark and flat field corrected with twilight sky flats, followed by hot and bad pixel correction. As with the $L_P$ data, the solution provided by \citet{Service:2016} was used to correct the NIRC2 Narrow camera geometric distortion. Finally, all the images were rotated to put north up. 

An additional step required for the $M_S$-band data that is not as critical for the other datasets is the background subtraction. Since the thermal background at 5$\mu$m is large and highly time variable, rather than median combine, or high pass filter to remove the background we adopted the least-squares sky subtraction algorithm proposed in \citet{Galicher:2011}. For each point in the dither pattern, the algorithm uses the images where the star is in one of the other three positions to construct a reference library. We used a ring centered on the star to estimate the thermal background in each image, with an inner annulus of 24 pixels and an outer annulus of 240 pixels. The final calibration step involved aligning the background corrected PSFs. Since the core of the PSF is saturated in the data, we aligned the data using two different methods, a) fitting a 2D Gaussian to the wings of the stellar PSF to estimate the center of the star and then shifting the PSF to a pre-determined pixel value to align all the images and b) using the rotation symmetry of the PSF using the method described in \citet{Morzinski:2015}. To compare the two methods, we calculated the residuals between images aligned using the methods and compared the noise in the residuals and found them to be similar and chose to go with the 2D Gaussian which is computationally faster.

The procedure used for the PSF subtraction for the $M_S$ data was similar to the $L_P$ data. The planet is not detected in each of the individual half-night datasets, requiring a combination of all four half-nights to increase the signal to noise ratio to detect the planet flux. To correctly combine the planet flux across the multiple epochs, we adjusted the PA to account for the astrophysical motion of the planet around the star, for which we used the best fitting orbit presented in \citet{DeRosa:2015}. In the month between the first and last dataset, the planet rotated $\sim$0.48 degrees or $\sim$0.4~pixel, which is a sufficiently large correction that it must be included in the data reduction. Each night's data was reduced individually to generate 603 PSF subtracted images. These images were then combined by dividing each image into 13 annuli which were combined using a weighted mean, where the weights are the inverse variance in each annulus. As seen in Figure~\ref{fig:gpiesK1K2LpMs}, we detect the planet signal at $\sim$2--3~sigma. To confirm that we are detecting the planet, we rotated the data to match the PA value of the $L_P$ epoch to find that the flux peak in the $M_S$-band matches the location of the planet in $L_P$. We measured a star to planet contrast of 11.5 mag using the same procedure as described for the $L_P$ data. We injected 25 fake PSFs that were scaled to match the contrast measured for the planet and detected the fakes at the same contrast as the planet. The final magnitude of the planet-star contrast in the $M_S$ is 11.5$\pm$0.5 mag.

\section{Results}
\label{sec:res}

To estimate stellar parameters of 51~Eri~A, \citet{Macintosh:2015} made use of Two Micron All-Sky Survey photometry \citep[2MASS;][]{Cutri:2003, Skrutskie:2006hla}. However, the $J$ and $H$-band photometry for the star are flagged as `E', indicating that the photometry is of the poorest quality and potentially unreliable (as compared to an `A' flag for the the $K$-band photometry). Further, the study used photometry taken with the \textit{Wide-field Infrared Survey Explorer} (WISE; \citealp{Wright:2010in}) in the $W1$ filter ($\lambda_\text{eff}$=3.35~$\mu$m, $\Delta\lambda$=1.11~$\mu$m) as an approximation for the $L_P$-band magnitude of the primary star. The photometry for 51~Eri~A in $W1$, from the AllWISE catalog \citep{Cutri:2013}, has large errors and contributes to more than half the error budget of the final planet photometry. In this study, we thus chose to re-observe the star in the $JHK_S$ filters and fit all the available photometry to estimate the photometry in filters where no calibrated stellar data exists.

\subsection{Revised Stellar Photometry at $J$,$H$,$K_S$}
\label{sec:sed}

The 2MASS near-IR colors of 51~Eri~A were compared to empirical colors for young F0 stars taken from \citet{Kenyon:1995}, where an F0IV star should have a $J-H$ = 0.13 mag and $H-K$ = 0.03 mag. The colors of 51~Eri~A estimated using the 2MASS photometry are however discrepant, with $J-H$ = $-0.03\pm0.08$, and $H-K$ = $0.23\pm0.08$ mag. The discrepant near-IR colors combined with poor quality flags suggest that the published photometry is potentially incorrect.

We observed the star 51~Eri~A using the 6.5-m MMT on Mt. Hopkins with the ARIES instrument \citep{McCarthy:1998} on 2016 Feb 28 UT under photometric conditions. We obtained data in the MKO $JHK_S$ broadband filters \citep{Tokunaga:2002}, for a total of 3.4 minutes in each filter. To flux calibrate these observations, we observed a photometric standard star at a similar airmass as 51~Eri~A, HR~1552 \citep{Carter:1990}. The raw images for both targets were processed through a standard near-IR reduction pipeline, performing dark current subtraction, flat field calibration, and bad pixels correction. Aperture photometry was performed on both targets, with the curve of growth used to select an aperture which minimized the error on the measured flux. The measured brightness of 51~Eri~A is presented in Table~\ref{tab:props}. 

Converting the MKO $K_S$-band measurement into the 2MASS system using empirical relations\footnote{\url{http://www.astro.caltech.edu/~jmc/2mass/v3/transformations/}} yields $K_{S, \text{2MASS}} = 4.551\pm0.032$~mag, which is within 1-$\sigma$ of the published 2MASS photometry. Furthermore, the $J-H$ and $H-K$ colors estimated from the revised photometry are $0.128\pm0.037$~mag and $0.016\pm0.039$~mag which are consistent with the empirical expectations.

\begin{figure}[!ht]
\centering
\includegraphics[width=\columnwidth]{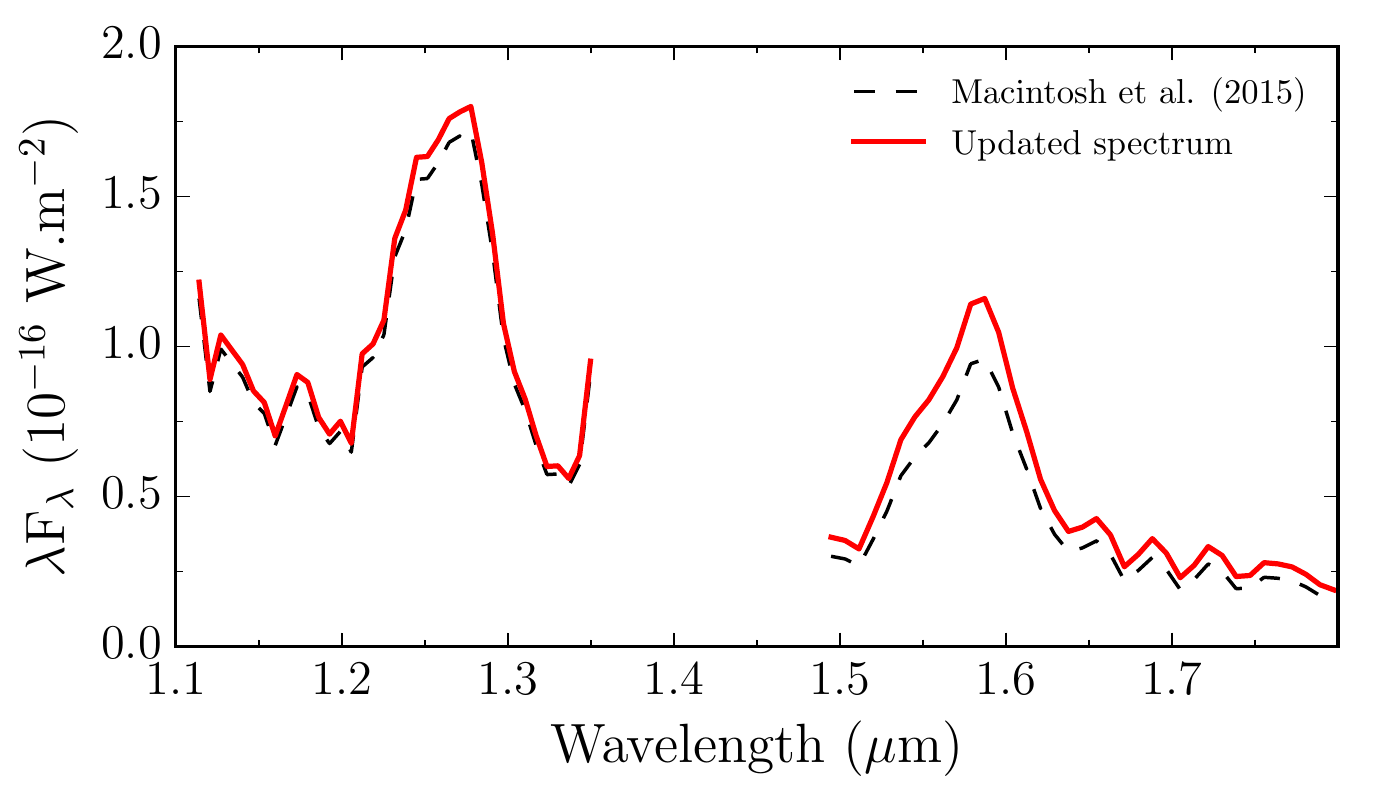}
\caption{ \label{fig:oldvnew} A comparison of the $JH$ spectra of 51~Eri~b using the literature 2MASS values against the new photometry measured in this study. The updated photometry increases the planet flux by $\sim$10\% in $J$ and $\sim$15\% in the $H$-band. The updated stellar photometry is used in the remainder of this study. However the final stellar spectrum used to correct the planet spectrum does not depend on individual filter photometry, as in \citet{Macintosh:2015} and shown in this plot, but is generated by modeling the full stellar SED prior to converting the planet spectra from contrast to flux units. }
\end{figure}

The published 51~Eri~b spectrum in \citet{Macintosh:2015} was calibrated using the Pickles stellar models \citep{Pickles:1998} to estimate the spectrum of the primary, where each band was scaled using the published 2MASS photometry. In Figure \ref{fig:oldvnew} we present a comparison between the published spectrum and one scaled using the new MKO photometry, using the same stellar models. The revised photometry scales the planet spectrum higher by $\sim$10\% in the $J$-band and $\sim$15\% in the H-band, which is significant given the high SNR of the $H$-band data.

\subsection{Fitting the Spectral Energy Distribution of 51~Eri~A }

\begin{figure}
\centering
\includegraphics[width=\columnwidth]{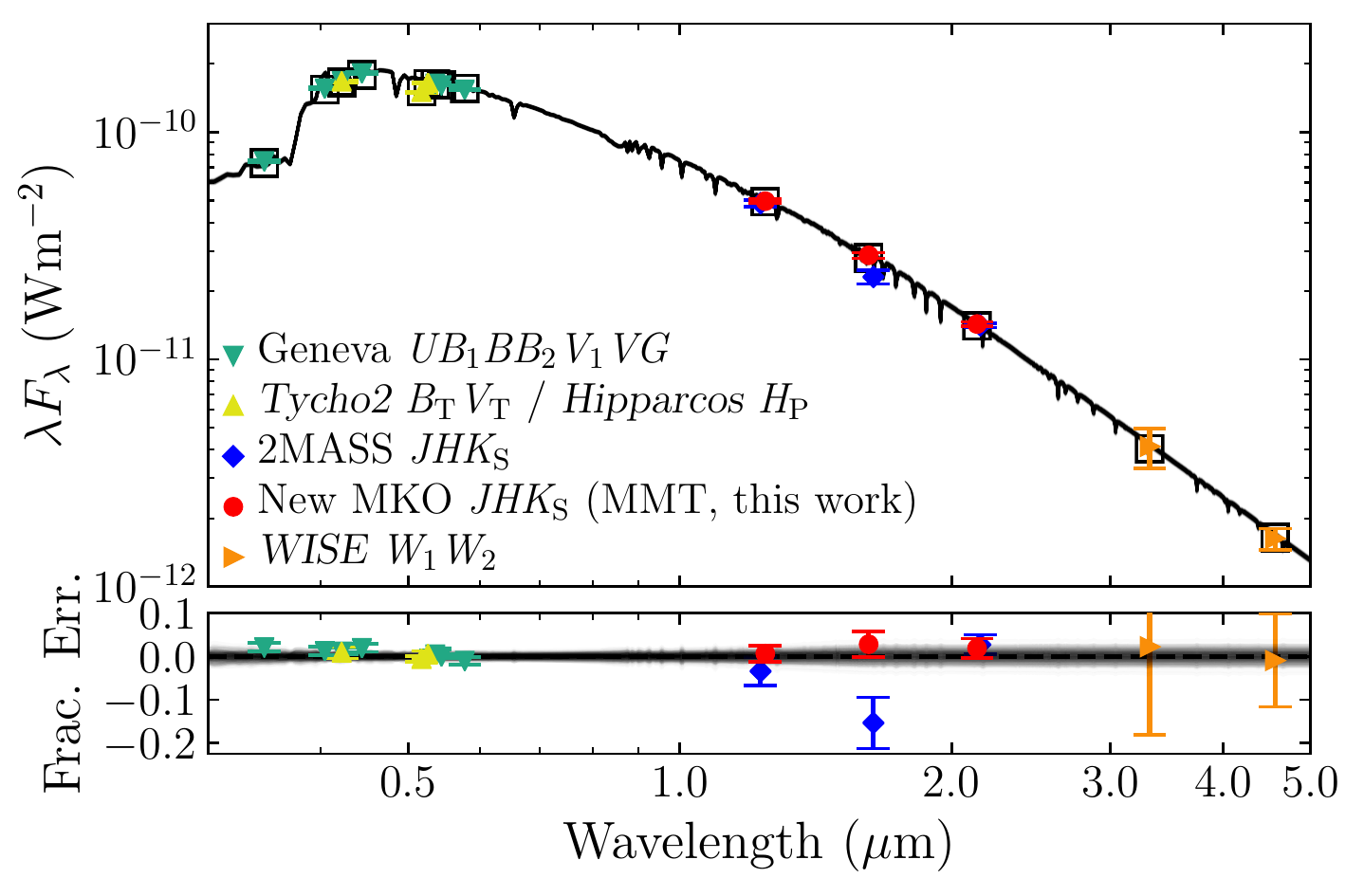}
\caption{ \label{fig:sed_A} (top panel): Photometry of 51~Eri~A from the literature, and from the results presented in this study (filled symbols). One hundred models were randomly selected from the MCMC search, and are plotted (translucent black curves). For each model, the synthetic magnitude was calculated for each filter. The median value for each filter is shown as an open square. The 2MASS photometry points are plotted to illustrate the offset relative to the new MKO measurements, and are not included in the fit. For the plotted Geneva photometry, we computed the Geneva $V$-band photometry using the best fit spectrum and then used the color relations to calculate the photometry in the remaining filters. (bottom panel): The fractional residuals relative to the median model.}
\end{figure}

To mitigate the effects of incorrect photometry, rather than scale the spectrum in pieces using the relevant broadband photometry, we decided to fit the full SED of 51~Eri~A using literature photometry and colors, including Geneva $U,B_1,B,B_2,V_1,V,G$ \citep{Rufener:1988}, \textit{Tycho2} $B_T,V_T$ / \textit{Hipparcos} $H_P$ \citep{Hog:2000,ESA:1997}, MKO $JHK_S$ (this work), and \textit{WISE} $W1,W2$ \citep{Cutri:2013} measurements. We made use of the Geneva color relations as constraints to the full SED fit since the published Geneva $V$ magnitude, which anchors the colors to estimate the remaining photometry, appears to be offset by $\sim$5\% when compared to the \textit{Tycho2} photometry. The \emph{WISE} $W2$ photometry was corrected using the \citet{Cotten:2016} relation for bright stars. We combine the photometry with model stellar atmospheres from the {\textsc BT-NextGen} grid\footnote{\url{https://phoenix.ens-lyon.fr/Grids/BT-NextGen/SPECTRA/}} \citep{Allard:2012fp}, we estimated the stellar spectrum using a five parameter MCMC grid search. The best fit atmosphere was found with $T_\text{eff} = 7331\pm30$ K, $\log g = 3.95\pm0.04$, [$M/H$] = $-0.12\pm0.06$, and a stellar radius, $R = 1.45\pm0.02$~$R_\odot$ (assuming a parallax of $33.98\pm0.34$~mas; \citealp{vanLeeuwen:2007}). No correction for extinction is performed as the extinction in the direction of 51~Eri is negligible ($A_V=0.00$; \citealp{Guarinos:1992um}). These values are consistent with previous literature estimates \citep[e.g.][]{Koleva:2012}. The final SED of 51~Eri~A is shown in Figure~\ref{fig:sed_A}, which highlights the significantly discrepant 2MASS $JH$-band photometry that was used previously to calibrate the spectrum of 51~Eri~b. We extracted MKO $K$, NIRC2 $L_P$ and $M_S$ photometry from the SED fit using the filter response functions presented in \citet{Tokunaga:2002}, see Table~\ref{tab:props}.

\subsubsection{Confirming the stellar $L_P$ photometry}
51~Eri~b emits a substantial amount of flux in the mid-IR and $L_P$ photometry in \citet{Macintosh:2015} was used to constrain the effective temperature of the planet. There exists no $L_P$ flux measurement for the star and thus they used the $W1$ magnitude reported in the AllWISE catalog ($W1 = 4.543\pm0.210$; \citealp{Cutri:2013}), and assumed a color of $W1 - L_P=0$ based on the F0IV spectral type of 51~Eri \citep{Abt:1995eo}. The $L_P$ photometry we estimated via the SED fits for 51~Eri is $L_P$ = $4.604\pm0.014$ mag, which is consistent with the value reported in \citet{Macintosh:2015} (4.52$\pm$0.21 mag) but with significantly smaller uncertainties.

As a final check for consistency, the 2MASS $K_S$ magnitude of 51~Eri ($K_\text{S, 2MASS} = 4.537\pm0.024$) was used instead as a starting point. The $K_S-L_P$ color for early F-type dwarfs and subgiants was estimated by folding model stellar spectra ($7200 \le T_\text{eff}/\text{K} \le 7400$, $4.0 \le \log g \le 4.5$, $[M/H]=0$) from the \textsc{BT-Settl} model grid through the relative spectral response of the 2MASS $K_S$ \citep{Cohen:2003gg} and NIRC2 $L_P$ filters. Over this range of temperatures and surface gravities, the color was calculated as $K_S-L_P = -0.001\pm0.001$. In order to realistically assess the uncertainties on this color, the near to thermal-IR spectra of F-type dwarfs and subgiants within the IRTF library \citep{Rayner:2009ki} were processed in the same fashion, resulting in a $K_S-L_P = 0.014\pm0.055$. A color of $K_S-L_P = -0.001\pm0.055$ was adopted based on the color calculated from the model grid, and the uncertainty calculated from the empirical IRTF spectra. This color, combined with the $K_{S, \text{2MASS}}$ magnitude of 51~Eri, gives an $L_P$ apparent magnitude of $4.538\pm0.060$. Each estimate for the stellar $L_P$ magnitude are within 1-$\sigma$ of each other, and thus we adopt the value derived from the SED fit i.e. $L_P$ = $4.604\pm0.014$ mag.

\subsection{51~Eri~b Spectral Energy Distribution}

\begin{figure}
\centering
\includegraphics[width=\linewidth]{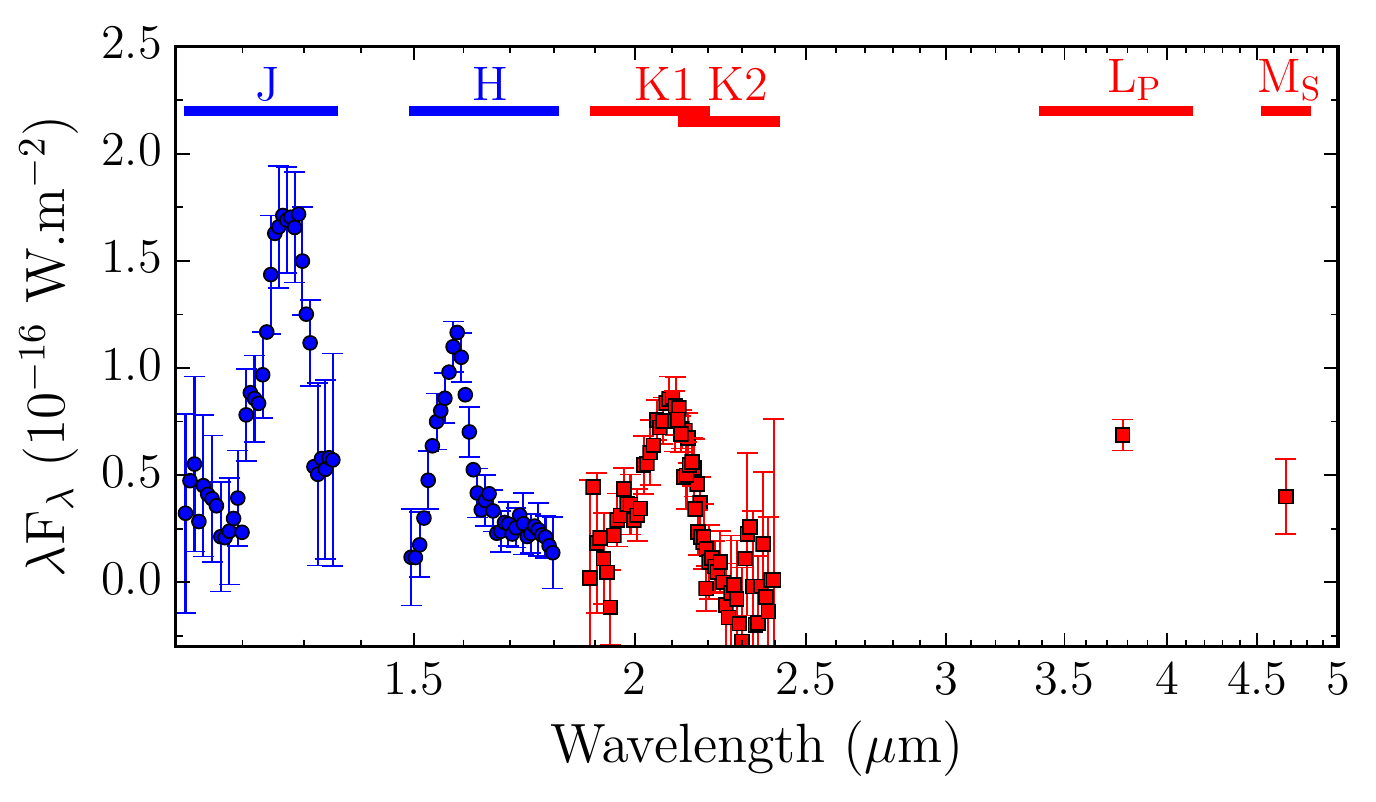}
\caption{ \label{fig:sed_b} Final spectral energy distribution of the directly imaged exoplanet 51~Eri~b. The new $K1$ and $K2$ GPI spectra along with the updated $L_P$ and new $M_S$ photometry are shown with red squares. The GPI $J$ and $H$ spectra, updated to account for the revised stellar flux, from the discovery paper \citep{Macintosh:2015} are plotted with blue circles. The filter extent is shown with the horizontal line over each band. To reduce crowding in the spectra, the errors for one out of every two data points are plotted.}
\end{figure}

We present the final spectral energy distribution of the planet 51~Eri~b in Figure~\ref{fig:sed_b} and use it to analyze the system properties in the following sections. Using the stellar SED estimated earlier, we have updated the $J$ and $H$ spectra that were published in \citet{Macintosh:2015}. In Table~\ref{tab:props}, we present the properties of the system, including updated MKO $JHK$ and NIRC2 $L_{P}M_{S}$ photometry for both the star and the planet. A future study will refine the orbital solution presented in \citet{DeRosa:2015}.

\begin{deluxetable}{lll}
\tablecaption{System properties. \label{tab:props}}
\tablewidth{0pt}
\tablehead{
\colhead{Property}	& \colhead{51~Eri~A} 	& \colhead{51~Eri~b}  
}
\startdata 
Distance (pc)   & \multicolumn{2}{c}{$29.43\pm0.29$\tablenotemark{a}}    \\
Age (Myr)       & \multicolumn{2}{c}{$26\pm3$\tablenotemark{b}}    \\
Spectral type     & F0IV              & T6.5$\pm$1.5 \\ 
$\log (L/L_\odot$)  & $0.85^{+0.06}_{-0.07}$ \tablenotemark{c}    & ${-5.83^{+0.15}_{-0.12}}$ to ${-5.93^{+0.19}_{-0.14}}$\tablenotemark{d} \\
$T_\text{eff}$       & 7331$\pm30$~K\tablenotemark{e}   & 605--737~K\tablenotemark{d} \\
$\log g$        & 3.95$\pm$0.04\tablenotemark{e}   &   3.5--4.0\tablenotemark{d} \\
$J_\text{MKO}$  & 4.690$\pm$0.020\tablenotemark{d}     & 19.04$\pm$0.40\tablenotemark{d,f} \\
$H_\text{MKO}$  & 4.562$\pm$0.031\tablenotemark{d}     & 18.99$\pm$0.21\tablenotemark{d} \\
$K_{S, MKO}$    & 4.546$\pm$0.024\tablenotemark{d}     & 18.49$\pm$0.19\tablenotemark{d} \\
$K_\text{MKO}$  & 4.600$\pm$0.024\tablenotemark{e}     & 18.67$\pm$0.19\tablenotemark{d} \\
$L_P$           & 4.604$\pm$0.014\tablenotemark{e}     & 16.20$\pm$0.11\tablenotemark{d,g} \\
$M_S$           & 4.602$\pm$0.014\tablenotemark{e}     & 16.1$\pm$0.5\tablenotemark{d} \\
\enddata
\tablenotetext{a}{\textit{Hipparcos} catalog \citep{vanLeeuwen:2007}}
\tablenotetext{b}{\citet{Nielsen:2016}}
\tablenotetext{c}{\citet{Macintosh:2015} using hot-start predictions.}
\tablenotetext{d}{This work}
\tablenotetext{e}{Stellar photometry estimated using SED fit}
\tablenotetext{f}{Distance Modulus = 2.34$\pm$0.02~mag}
\tablenotetext{g}{Weighted mean of the two $L_P$ observations}
\end{deluxetable}

\section{Analysis}
\label{sec:ana}

\begin{figure*}
\centering
\includegraphics[width=\linewidth]{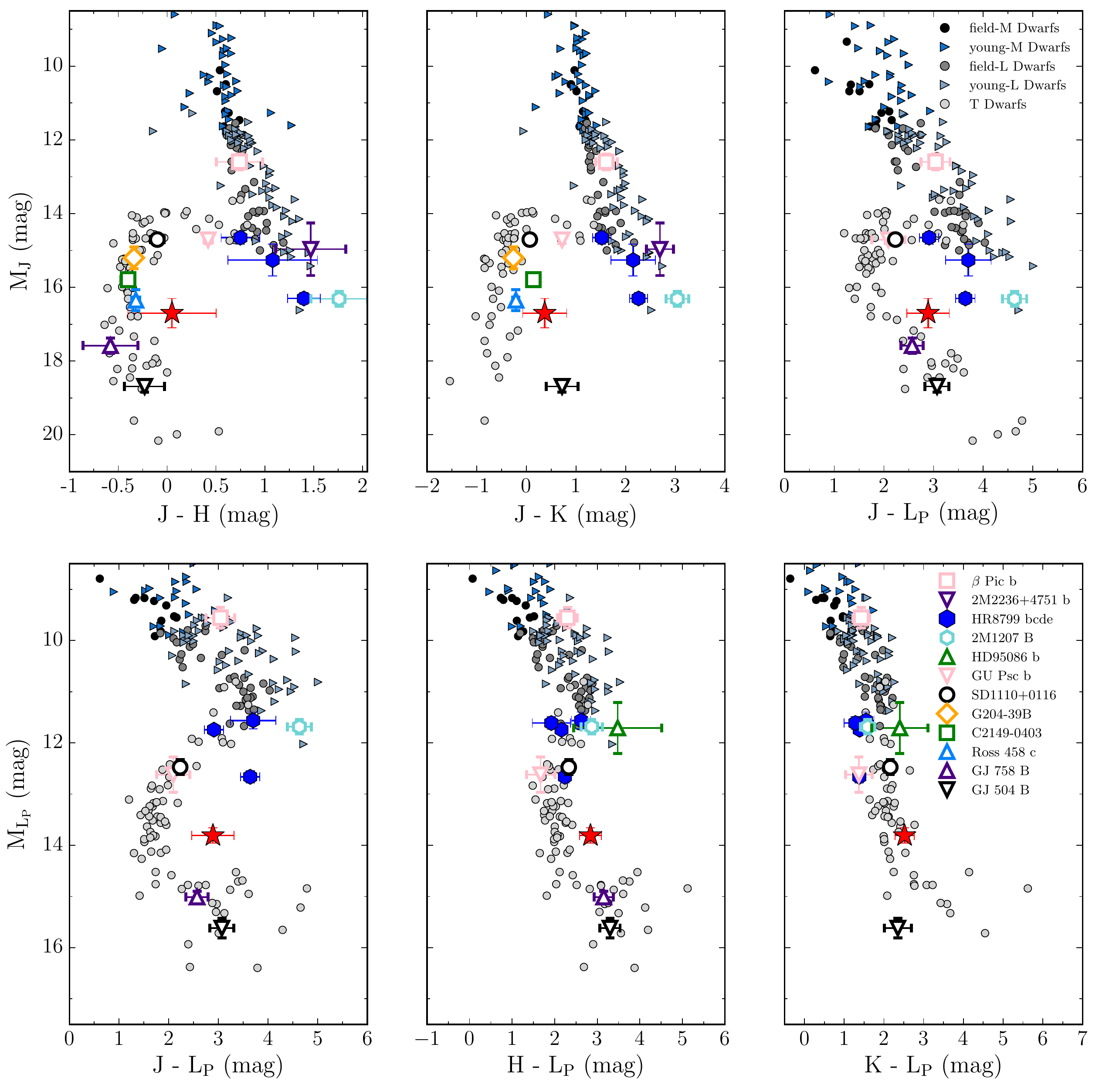}
\caption{ \label{fig:CMD} The brown dwarf and imaged exoplanet color magnitude diagram. 51~Eri~b is shown with the red star. The colors of 51~Eri~b place it among late T-dwarfs, where it is redder than most comparable temperature brown dwarfs likely indicative of greater cloud opacity in the atmosphere. The photometry for the field M-dwarfs (black circles), young M-dwarfs (blue triangles), field L-dwarfs (dark gray circles), young L-dwarfs (light blue triangles), and T-dwarfs (light gray circles) is taken from the compilation of \citet{Dupuy:2012, Liu:2016}. We used a linear fit to convert \emph{WISE} $W1$ photometry to $L_P$, similar to what was done in \citet{Macintosh:2015}. The photometry for the directly imaged planets and young brown dwarfs were taken from \citet{Males:2014, Bonnefoy:2014, Bowler:2017, Marois:2010b, Chauvin:2005a, Rameau:2013, Naud:2014, Leggett:2007, Delorme:2017, Goldman:2010, Janson:2011, Kuzuhara:2013}.}
\end{figure*}

\subsection{Comparison against field brown dwarfs}
\label{ssec:comp}

\begin{figure}
\centering
\includegraphics[width=\columnwidth]{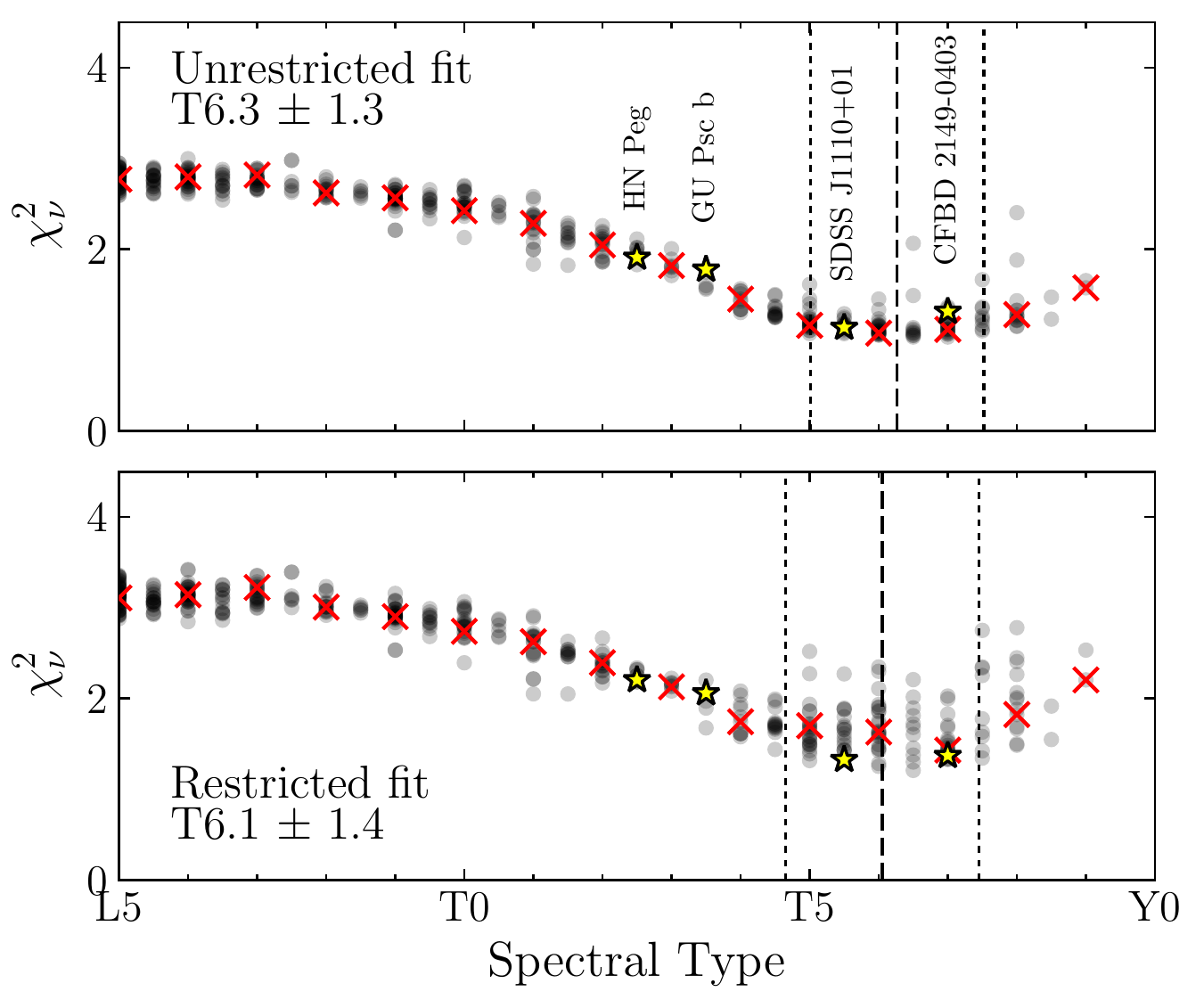}
\caption{Comparison of L5 to T9 field (gray circles) and young (yellow stars) brown dwarf $JHK$ spectra to 51~Eri~b using the reduced $\chi^2$. The standard brown dwarf for each spectral bin is plotted with a red cross \citep{Kirkpatrick:2010, Burgasser:2006a, Cushing:2011}. The dashed and dotted vertical lines give the best fitting spectral type, and corresponding uncertainty. (Top) Each spectral band of the comparison was allowed to float to find the lowest chi-square while fitting the planet spectrum. (Bottom) The spectrum was allowed to float up/down in flux, but was penalized by the spot ratio uncertainty in each respective band.  \label{fig:fits_all} }
\end{figure}

We plot a series of color magnitude diagrams (CMD) for ultracool objects in Figure~\ref{fig:CMD}, and compare the photometry of field M, L, and T dwarfs and young brown dwarfs and imaged companions to that of 51~Eri~b (red star). The colors of 51~Eri~b seems to match the phase space of the late-T dwarfs. To classify the spectral type of 51~Eri~b, we do a chi-square comparison of the GPI $JHK1K2$ spectrum of 51~Eri~b to a library of brown dwarf spectra compiled from the IRTF \citep{Cushing:2005}, SpeX \citep{Burgasser:2014}, and Montreal \citep[e.g.][]{Gagne:2015a, Robert:2016} Spectral Libraries. Only a small sub-sample of the brown dwarfs have corresponding mid-IR photometry and thus we choose to restrict our comparison to the near-IR. The spectra within the library were convolved with a Gaussian kernel to match the spectral resolution of GPI. 

To compute the chi-square between the spectrum of 51~Eri~b and the objects within the library, we use two different equations. The first method permits each individual filter spectrum to vary freely (unrestricted fit). In the unrestricted fit, we compute the $\chi^2$ statistic for the $j^\text{th}$ object within the library as

\begin{equation}
\chi^2_j = \sum_{i=1}^4 \left({\bm S}_i - \alpha_{i,j}{\bm F}_{i,j}\right)^T{\bm C}_i^{-1}\left({\bm S}_i - \alpha_{i,j}{\bm F}_{i,j}\right),
\label{chi-square_unrestricted}
\end{equation}

where ${\bm S}_i$ is the spectrum of the planet, ${\bm C}_i$ is the covariance matrix calculated in Section \ref{app:cov2}, and ${\bm F}_{i,j}$ is the spectrum of the $j^\text{th}$ comparison brown dwarf, all for the $i^\text{th}$ filter. For each object, the scale factor $\alpha_{i, j}$ that minimizes $\chi^2$ is found using a downhill simplex minimization algorithm. In this method the scale factor for each object, $\alpha_{i, j}$, is allowed to vary between the four filters ($JHK1K2$). This is equivalent to allowing the near-IR colors to vary freely up and down in order to better fit the object \citep[e.g.][]{Burningham:2011}. 

In the second method the individual filter spectra are still allowed to vary, only within the satellite spot brightness ratio uncertainty (restricted fit), thereby restricting the scale factor for each filter. For the restricted fit the scale factor is split into two components. The first, $\alpha_j$, is independent of filter, and accounts for the bulk of the difference in flux between 51~Eri~b and the comparison object due to differing distances and radii. The second, $\beta_{i,j}$, is a filter-dependent factor that accounts for uncertainties in the satellite spot ratios given in \citet{Maire:2014}. Equation \ref{chi-square_unrestricted} is modified to include an additional cost term restricting the possible values of $\beta_{i,j}$, 

\begin{multline}
\chi^2_j = \sum_{i=1}^4 \left[\left({\bm S}_i - \alpha_j\beta_{i,j}{\bm F}_{i,j}\right)^T{\bm C}_i^{-1}\left({\bm S}_i - \alpha_j\beta_{i,j}{\bm F}_{i,j}\right) \right.\\ + \left. N_i \left(\frac{\beta_{i,j}-1}{\sigma_{i}}\right)^2\right]
\label{chi-square}
\end{multline}

where $N_i$ is the number of spectral channels in the 51~Eri~b spectrum for the $i^\text{th}$ filter, and $\sigma_i$ is the uncertainty on the satellite spot flux ratio given in \citet{Maire:2014} for the same filter. The second term in Equation \ref{chi-square} penalizes values of the scale factor, $\beta_{i,j}$, that are very different from the satellite spot uncertainty and thus increases the chi-square for objects significantly different from 51~Eri~b.

The spectral type of 51~Eri~b was estimated for both fits from the $\chi^2$ of the L5--T9 near-IR spectral standards \citep{Kirkpatrick:2010, Burgasser:2006a, Cushing:2011}. To compute the weighted mean and standard deviation of 51~Eri~b, we converted the spectral type to a numerical value for the standard brown dwarfs, i.e. L5 = 75, T5 = 85. Each numerical spectral type when compared to 51~Eri~b, is weighted according to the ratio of its $\chi^2$ to the minimum $\chi^2$ for all standards (e.g., \citealp{Burgasser:2010}), and the lowest value was adopted as the spectral type of 51~Eri~b. A systematic uncertainty of one half subtype was assumed for the standards. We find that the two estimates are consistent with one another i.e. T$6.3\pm1.3$ and $6.1\pm1.4$ for unrestricted and restricted fits, see Figure~\ref{fig:fits_all}. We adopt a spectral type for 51~Eri~b of T$6.5\pm1.5$ from the unrestricted fit, rounded to the nearest half subtype.

\begin{figure}
\centering
\includegraphics[width=\columnwidth]{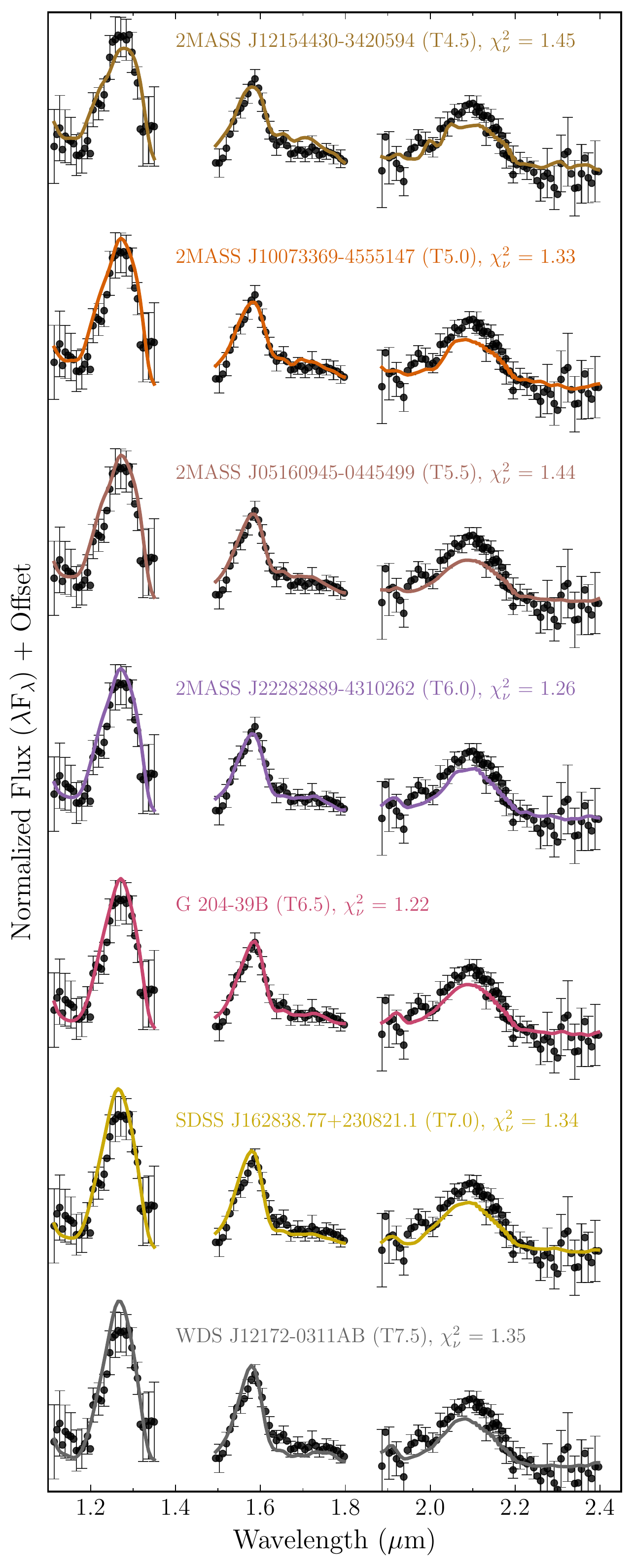}
\caption{ \label{fig:fit_mid_Ts} Comparing the spectra, using the restricted fit, of the best fitting T4.5 to T7.5 field brown dwarfs to 51~Eri~b. The spectra shown in this figure are a subset of the data plotted in Figure \ref{fig:fits_all}. The brown dwarf spectral fits plotted here use the restricted chi-square equations presented in Equation~\ref{chi-square}. The T4.5 and T5.0 spectra are from \citet{Looper:2007}, the T5.5 is from \citet{Burgasser:2008}, the T6.0 is from \citet{Burgasser:2004}, the T6.5 and T7.5 are from \citet{Burgasser:2006b}, and the T7.0 is from \citet{Dupuy:2012}.}
\end{figure}

The best-fit object for both the unrestricted and restricted fits was G~204-39~B (SDSS~J175805.46+463311.9; $\chi^2_{\nu} = 1.033$ and $1.209$), a T6.5 brown dwarf common proper motion companion to the nearby M3 star G~204-39~A \citep{Faherty:2010}. G~204-39~B has marginally low surface gravity based on photometric ($\log g \approx 4.5$; \citealp{Knapp:2004}) and spectroscopic measurements ($\log g = 4.7$--$4.9$; \citealp{Burgasser:2006b}), indicative of it being younger than the field population. While the binary system is not thought to be a member of any known young moving group \citep{Gagne:2014}, the stellar primary can be used to provide a constraint on the age of the system. Combining the X-ray and chromospheric activity indicators for the M dwarf primary, and a comparison of the luminosity of the secondary with evolutionary models, \citet{Faherty:2010} adopt an age of 0.5--1.5\,Gyr for the system. 51~Eri~b is redder than the spectrum of G~204-39~B (Figure \ref{fig:fit_mid_Ts}), especially in terms of the $H-K$ color, which is a photometric diagnostic of low surface gravity among T-dwarfs \citep[e.g.,][]{Knapp:2004}. This is consistent with the younger age of 51~Eri~b, and the most likely cause for this is that it has lower surface gravity than that of G~204-39~B.

Additional good matches to the 51~Eri~b spectrum include 2MASS~J22282889--4310262 (2M~2228--43, $\chi^2_{\nu} = 1.07$ and $1.26$ for the two fits) and 2MASS~J10073369--4555147 (2M~1007--45, $\chi^2_{\nu} = 1.07$ and $1.33$). 2M~2228--43 a well-studied T6 brown dwarf that exhibits spectrophotometric variability in multiple wavelengths indicative of patchy clouds in the photosphere \citep{Buenzli:2012,Yang:2016}. 2M~1007--45 is a T5 brown dwarf at a distance of $17\pm2$~pc \citep{Smart:2013}. It was identified by \citet{Looper:2007} as a low surface gravity object based on its H$_2$O$-J$ vs $K/H$ spectral ratios defined in \citet{Burgasser:2006b}; comparisons against solar-metallicity models imply an age of between 200 and 400~Myrs \citep{Looper:2007}.

The best fit object for each spectral type between spectral types T4.5 and T7.5 using the restricted fit are plotted in Figure~\ref{fig:fit_mid_Ts}. While the quality of the fits were generally good, none of the objects were able to provide a good match across all of the bands simultaneously, being too luminous in either the $J$ or $K$-bands. Differences in surface gravity, effective temperature, and/or metallicity could be the cause \citep[e.g.,][]{Knapp:2004}. The poor fit to the color of 51~Eri~b is especially apparent in the CMDs plotted in Figure~\ref{fig:CMD}, with 51~Eri~b having unusually red near-IR colors relative to similar spectral type objects.

\subsection{Comparison against young brown dwarfs}

\begin{figure}
\centering
\includegraphics[width=\columnwidth]{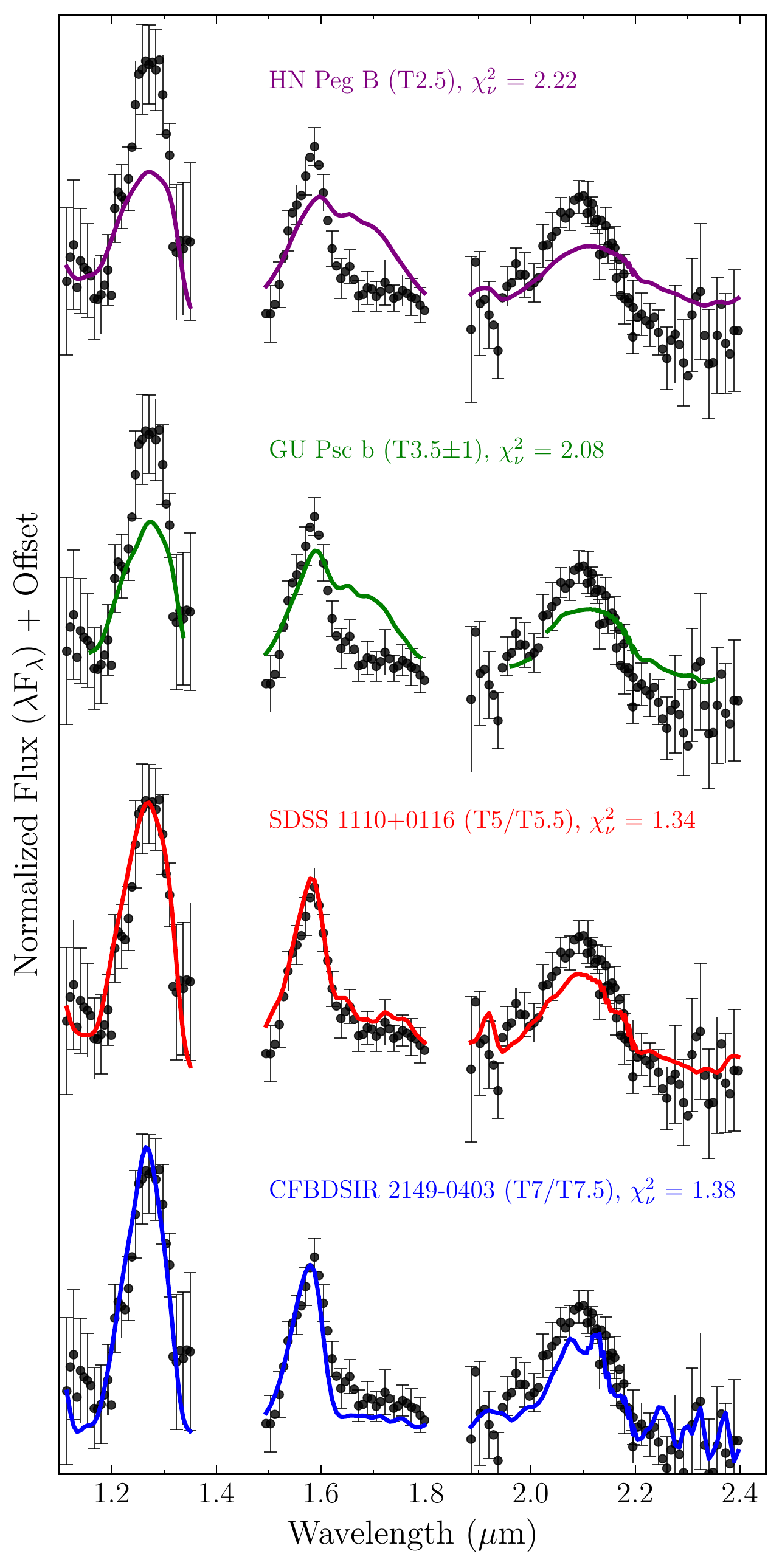}
\caption{ \label{fig:fits_yng_Ts} Comparing the spectra of known young T-dwarfs to that of 51~Eri~b. Similar to the field sequence, the fits presented here were computed using the restricted chi-square. From top to bottom, the four spectra were sourced from \citet{Luhman:2007}, \citet{Naud:2014}, \citet{Burgasser:2006b}, and \citet{Delorme:2012}. }
\end{figure}

Searches for young companions and moving group objects have resulted in detections of several tens to hundred of million year old L-type brown dwarf and planetary mass companions as well as the identification of L-dwarf sub-classes based on youth \citep[e.g.][]{Allers:2013, Filippazzo:2015, Faherty:2016, Liu:2016}. In comparison, there exist relatively few known (or suspected) young T-dwarf brown dwarfs. In Figure~\ref{fig:fits_yng_Ts}, we plot the known young T-dwarfs and compare them in a similar manner to what was done above for field brown dwarfs. The chi-square for the fits is not much better than what is seen for the field dwarfs which is likely due to the absence of young T-dwarfs of similar spectral type to 51~Eri~b. 

The brown dwarf SDSS~J1110+0116 with a spectral type of T5/T5.5 is the best fitting young comparison object. It has been identified as a \textit{bona fide} member of the AB~Doradus moving group and is thus young (110--130~Myr) and low mass (10--12\,M$_\text{Jup}$) \citep{Gagne:2015b}. The other young field object that closely matches the near-IR spectrum of 51~Eri~b is the T7 peculiar brown dwarf, CFBDSIR~J2149-0403 \citep{Delorme:2012}. CFBDSIR~J2149-0403 was originally suggested to be a member of of the AB~Doradus moving group, however \citet{Delorme:2017} find that the parallax and kinematics of the free-floating object rule out its membership to any known young moving group. However, despite the lack of proof of youth, medium resolution spectroscopy examining the equivalent width of the KI doublet at 1.25$\mu$m suggests that the object has low surface gravity, and is most likely a young planetary mass object (2--13~M$\text{Jup}$). An alternative solution is that it is a higher mass, 2--40~M$\text{Jup}$, brown dwarf with high metallicity. CFBDSIR~J2149-0403 shows stronger methane absorption features in the red end of the $H$-band spectrum as compared to 51~Eri~b. However, it is worth pointing out that while both young objects, SDSS~J1110+0116 and CFBDSIR~J2149-0403, are reasonable matches across the $J$ and $H$ spectra of 51~Eri~b, they appear to be under-luminous in the $K$-band. A likely reason for this is that 51~Eri~b is much younger than both the comparison companions and thus has the lowest surface gravity amongst the three objects \citep{Burgasser:2006b}. 

\subsection{A very red T6 or an L-T transition planet?}
\label{ssec:LT}

\begin{figure}
\centering
\includegraphics[width=\columnwidth]{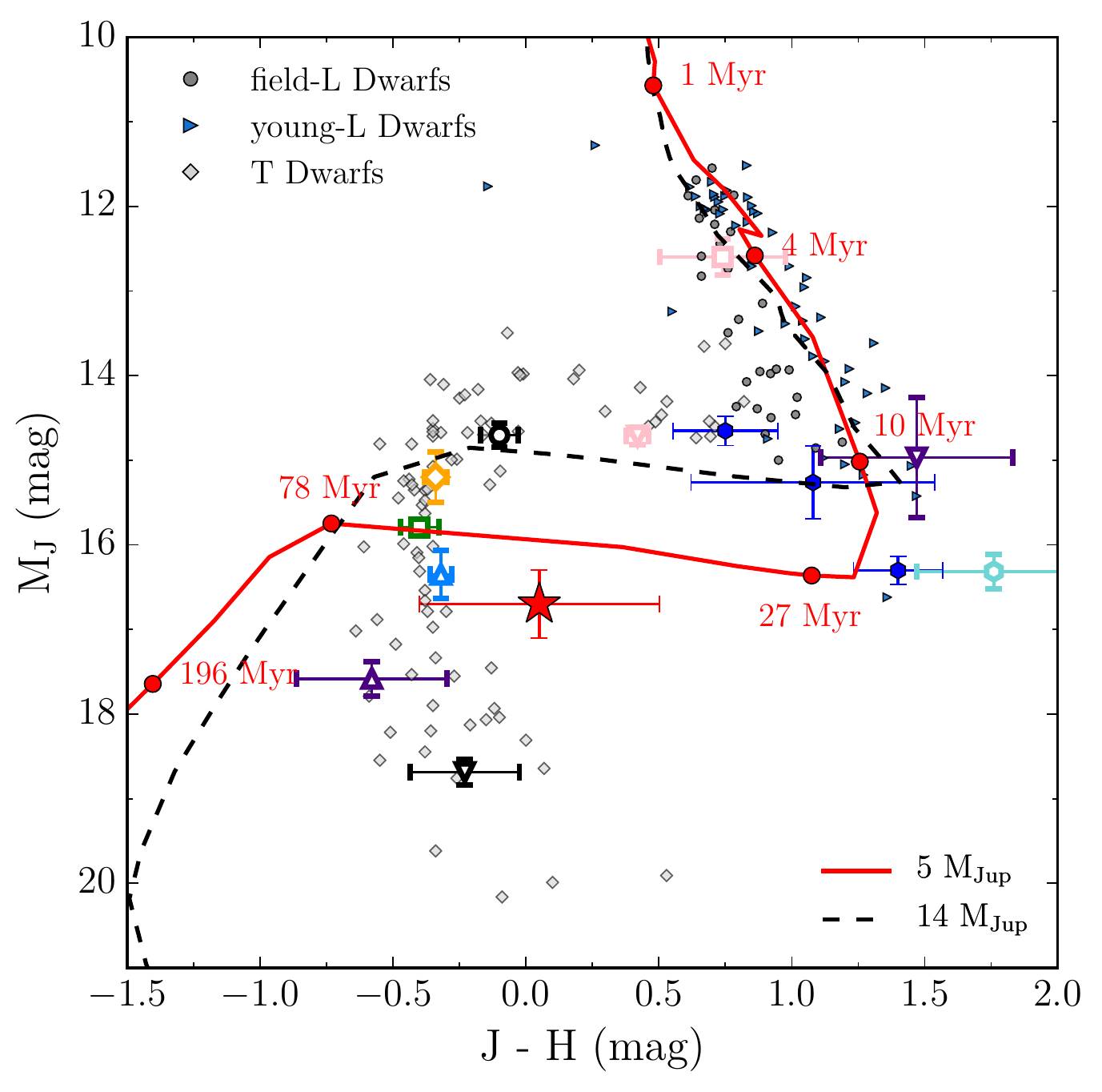}
\caption{ \label{fig:cmd_wtracks} The $J$ vs $J-H$ brown dwarf and imaged exoplanet color magnitude diagram reproduced from Figure~\ref{fig:CMD}. The photometry for 51~Eri~b is shown with the red star. Also plotted on the CMD are the evolutionary tracks for 5 and 14 M$_\text{Jup}$ objects \citep{Marley:2012}, with the solid red line and dashed black line respectively. The models assume a simple gravity dependence for the initiation of the transition. A few ages for the 5~M$_\text{Jup}$ track have been over plotted. The L-T transition for the 5 M$_\text{Jup}$ planet starts at approximately 900 K and 20 Myr, but for a lower mass planet such as 51~Eri~b will occur at younger ages. }
\end{figure}

Based on the position of 51~Eri~b in Figure~\ref{fig:CMD}, it appears that the trend of planetary mass objects having redder colors compared to the field, seen in young L-type brown dwarfs and planetary mass companions \citep{Faherty:2016, Liu:2016}, possibly continues for the T-type companions. Note that the $K-L_P$ CMD shows little reddening, which is natural if clouds are causing the effect. The effect of clouds is negligible in the $K$ and $L_P$ bands. Across both the near and mid-IR CMDs, 51~Eri~b is one of the reddest T type objects and within its spectral classification it has the reddest colors. This trend in the 51~Eri~b colors was originally noted in \citet{Macintosh:2015} where they compared the $L_P$ vs $H-L_P$ color for the planet and noted that it was clearly redder than the field. Rather than simply being redder than the field T-dwarfs due to the presence of clouds, we present a second possible interpretation for the red colors of 51~Eri~b, of the planet still undergoing the process of transitioning from L-type to T-type. This hypothesis assumes that the evolutionary track followed is gravity dependant with examples for higher mass objects shown in Figure~\ref{fig:cmd_wtracks}. In this scenario, 51~Eri~b transitions at fainter magnitudes than that seen for field L-T transition brown dwarfs and it has not yet completed its evolutionary transition to reach the blue colors typical of field, mid-T dwarfs. 

In Figure~\ref{fig:cmd_wtracks}, we re-plot the $J$ vs $J-H$ panel from the series of CMDs shown in Figure~\ref{fig:CMD}. In addition to the photometry of 51~Eri~b and the field and young brown dwarfs we also over-plot two low mass, 5 and 14 M$_\text{Jup}$, evolutionary model tracks (assuming hot-start conditions) from \citet{Saumon:2008, Marley:2012}. If the L-T transition is gravity dependent, as multiple lines of evidence now suggest \citep{Leggett:2008, Dupuy:2009, Stephens:2009}, then lower mass objects may turn blue at fainter absolute magnitudes than field objects. In Figure~\ref{fig:cmd_wtracks}, we show a simple model in which the L to T transition begins at 900 K at $\log g = 4$ (solid red line) instead of 1200 K at $\log g = 5.3$ (dashed black line). In the case of a $5 \text{M}_\text{Jup}$ planet the L to T transition begins and ends about 1 magnitude fainter in $J$ band than observed for the field population. Furthermore the congruence of the spectrum of SDSS~J1110+0116 with 51~Eri~b (Figure~\ref{fig:fits_yng_Ts}) is interesting as SDSS~J1110+0116 lies just short of the blue end of the field L to T transition, although it does so at an absolute magnitude just slightly fainter than the field transition magnitude. While these simple models explain the fainter absolute magnitudes of the transition, their colors are too blue and appear to miss the younger brown dwarf and free-floating planets. Similarly the models are too blue to match the T dwarf sequence. Clearly more sophisticated modeling of evolution through the L to T transition, accounting for inhomogeneous cloud cover and a gravity dependent transition mechanism as well as a range of initial conditions is required. Testing this hypothesis is difficult and would require knowledge of the true mass of the companion as well as the formation mechanism. If this hypothesis is true, then the only objects that are brighter on the CMD should be higher mass objects. There should not be any lower mass objects above and to the left of 51~Eri~b on the $J$ vs $J-H$ CMD shown in Figure~\ref{fig:cmd_wtracks}. 

\begin{deluxetable*}{cccccc}
\tablecolumns{6} 
\tablewidth{0pt} 
\tablecaption{Model grid parameters \label{tab:model_params}} 
\tablehead{ 
\colhead{Model} & \colhead{Effective} & \colhead{Surface Gravity} & \colhead{Metallicity} & \colhead{Cloud} & \colhead{Cloud Hole}\\
\colhead{Name}  & \colhead{Temperature (K)} & \colhead{[$\log g$] (dex)} & \colhead{[M/H] (dex)} & \colhead{Parameter (f$_\text{sed}$)} & \colhead{Fraction (\%)} }
\startdata 
Iron/Silicate Cloud Grid  & 600--1000 & 3.25    & 0.0          & 2          & 0--75 \\ 
Sulfide/Salt Cloud Grid & 450--900  & 3.5--5.0 & 0.0, 0.5, 1.0 & 1, 2, 3, 5 & -- \\
Cloudless Grid          & 450--900  & 3.5--5.0 & 0.0, 0.5, 1.0 & no cloud   & -- \\
\enddata 
\end{deluxetable*}

\section{Modeling the atmosphere of 51~Eri~\lowercase{b}}

For the purpose of modeling the complete SED of 51~Eri~b we made use of two updated atmospheric model grids from the same group, focusing on different parameter space (see Table~\ref{tab:model_params}). The first grid, described in \citet{Marley:1996, Marley:2002, Marley:2010} focused on the higher effective temperature atmospheres (L-dwarfs) and includes iron and silicates clouds in the atmosphere. The second grid, described in \citet{Morley:2012, Morley:2014} and \citet{Skemer:2016}, is designed for lower effective temperatures (T and Y dwarfs) and include salt and sulfide clouds in the atmosphere, which are expected to condense in the atmospheres of mid to late-T dwarfs. 

The methodology used to fit the models to the data is the same for both model grids. To fit the models to the data, we bin the model spectra to match the spectral resolution of the GPIES spectra across each of the $JHK1K2$ filters. For the photometry we integrated the model flux through the Keck/NIRC2 $L_P$ and $M_S$ filter profiles respectively. The estimation of the best fitting model is done by computing the chi-square value for each model in the grid compared to the data using Equation~\ref{chi-square}. We made use of the covariance matrices estimated for the four spectral channels described in the appendix and also included the variance for each of the two photometric data points to compute the chi-square statistic. Note that we use the restricted fit equation in the computation of the best fitting model. This equation permits each of individual filters to scale within the 1-$\sigma$ error of the satellite spot ratios. We also did the fitting without the scaling factor and found that the results are similar.  

As stated earlier in section~\ref{sec:covariance}, the use of the covariance affects the model fitting where the peak of the posterior distribution occurs at slightly cooler effective temperatures, consistent within the errors. Due to the high spectral correlation in the $J$-band (see Figure~\ref{fig:psi}), when using the covariance the best fitting models are not models that pass through the data but rather models that have lower flux in the $J$-band than the data. We present the specific modeling details in the following text.
\label{model}

\subsection{Iron and Silicates Cloud Models}

\begin{figure}[!ht]
\centering
\includegraphics[width=\columnwidth]{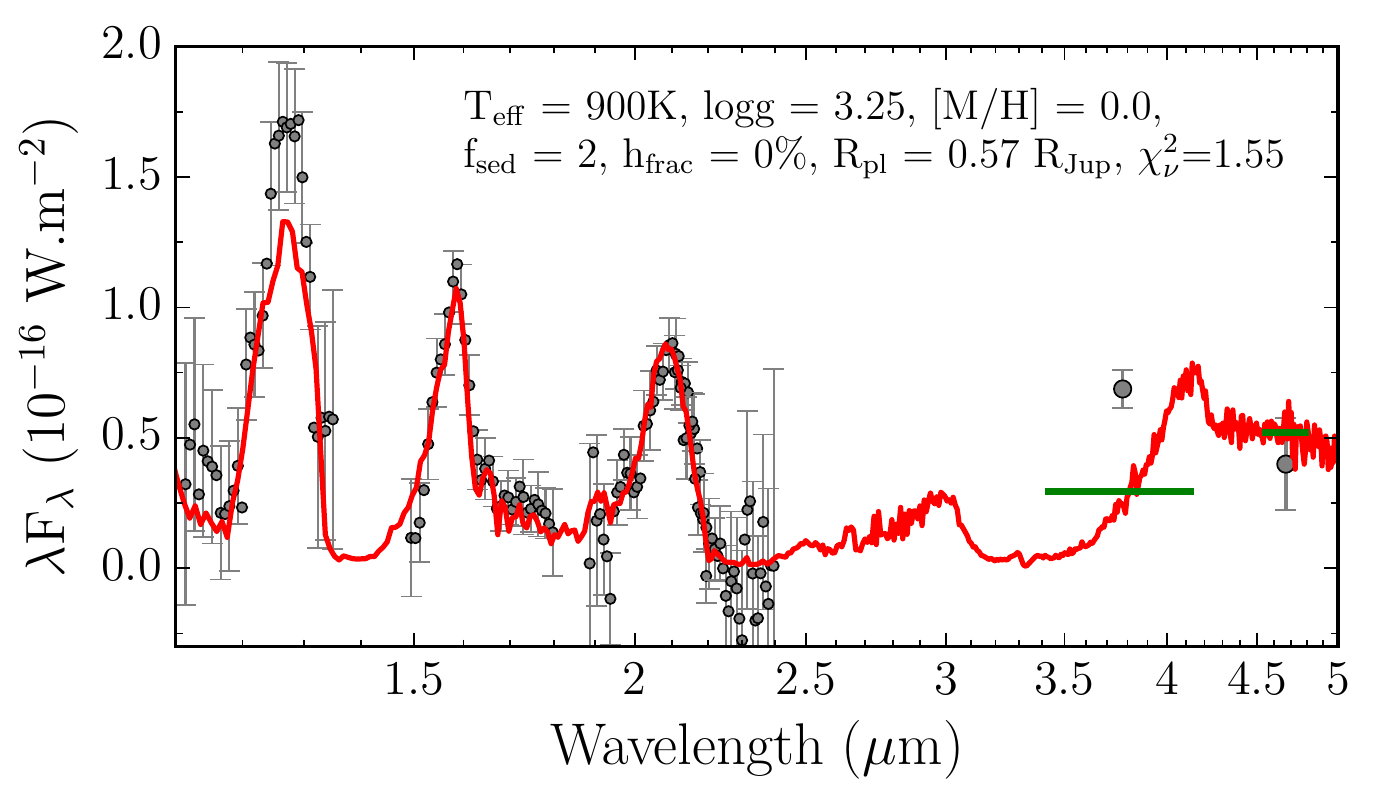}
\caption{ \label{fig:MM_modelfits} Spectral Energy Distribution of 51~Eri~b with the best fitting iron and silicates cloudy model. } 
\end{figure}

\begin{figure*}[!ht]
\centering
\includegraphics[width=12cm]{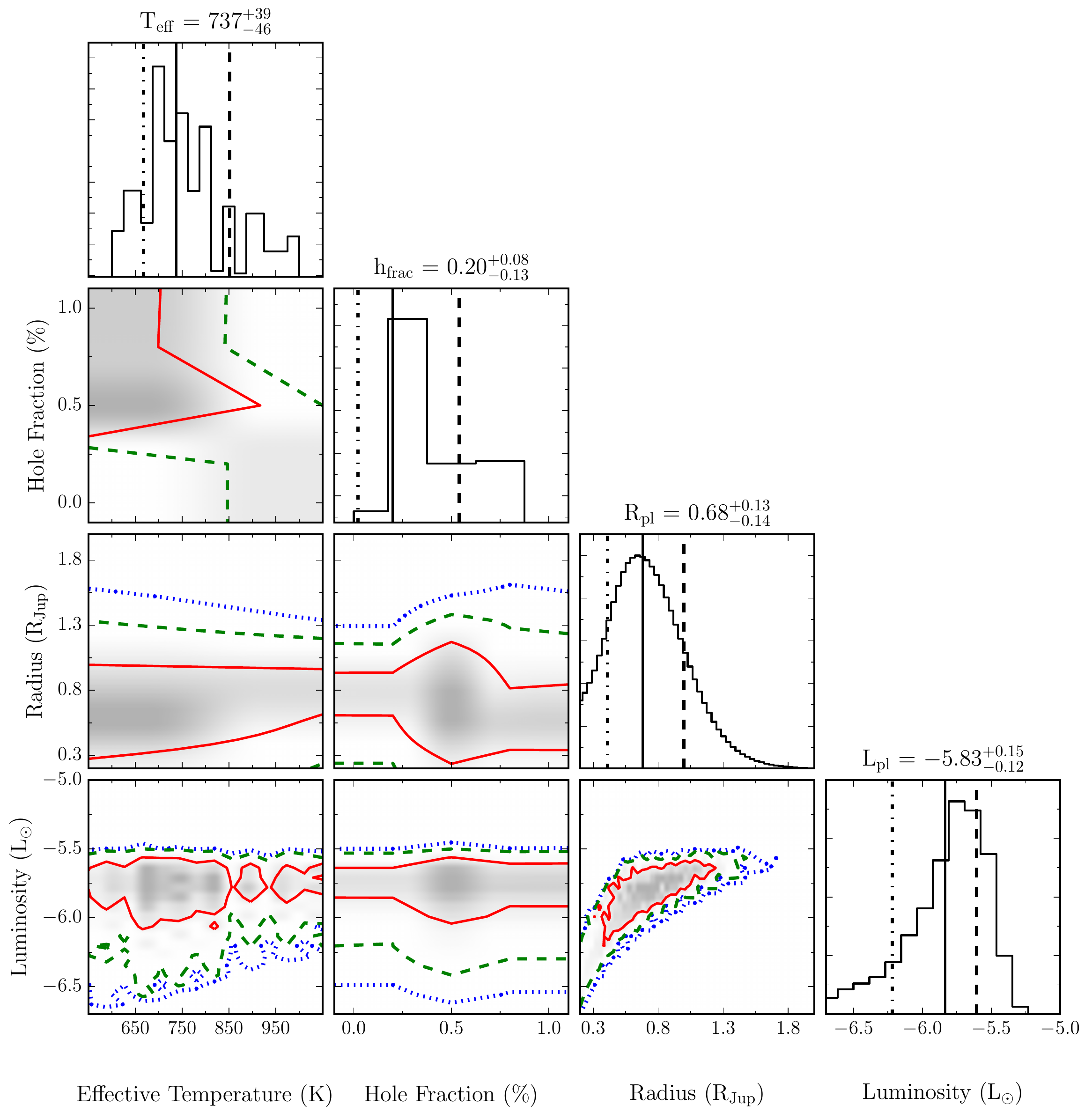}
\caption{ \label{fig:MM_posterior} Normalized posterior distributions for the iron and silicates model grid. The PDFs are for the parameters varied in our fit along with the inferred distribution of the luminosity of 51~Eri~b. The lines on the 1D histogram indicate the 16th, 50th, and 84th percentile values while those on the 2D histogram are the 1$\sigma$ (solid red), 2$\sigma$ (dashed green) and 3$\sigma$ (dotted blue) values of the distribution. The values printed above each histogram are the median value along with the 1-$\sigma$ error on it.}
\end{figure*}

In sec.~\ref{ssec:LT}, we suggested that 51~Eri~b, rather than having completely evolved to T-type, could be transitioning from L-to-T. In this scenario the cloud composition of the planetary atmosphere might still be influenced by the deep iron and silicates condensate grains and patchy cloud atmosphere. Therefore, we compared the planet SED to a grid of models with a fixed low surface gravity and solar metallicity, where the key variable is cloud hole fraction and the unique aspect of this grid is the presence of iron/silicate clouds in an atmosphere with clear indications of methane absorption. The clouds are modelled using the prescription presented in \citet{Ackerman:2001}, where cloud thickness is parameterized via an efficiency factor (f$_\text{sed}$). Where small values of f$_\text{sed}$ indicate atmospheres with thick clouds while large values of f$_\text{sed}$ are for atmospheres with large particles that rain out of the atmosphere leaving optically thinner clouds. As mentioned early the primary condensate species in this grid are iron, silicate, and corundum clouds, molecules that are expected to dominate clouds in L-dwarfs \citep{Saumon:2008, Stephens:2009}. At the L-T transition clouds are expected to be patchy, thus for each T$_\text{eff}$, the models went from fully cloudy i.e. f$_\text{sed}$ = 2 and 0\% holes to an atmosphere with f$_\text{sed}$ = 2 and 75\% holes (patchy clouds). The methodology used to calculated the flux emitted from the patchy cloud atmosphere include both cloud and cloud-free regions simultaneously in the atmosphere using a single, global temperature-pressure profile and are not created via a linear combination of two models as is sometimes done in the literature \citet{Marley:2010}. The iron and silicates cloud grid models use solar metallicity \citep{Lodders:2003}. The opacity database used for the absorbers are described in \citet{Freedman:2008}, including updated molecular line lists for ammonia and methane \citep{Yurchenko:2011, Yurchenko:2014}. The models span effective temperatures from 600K to 1000K for solar metallicity ([M/H] = 0.0) and low surface gravity ($\log g$ = 3.25, 3.50) (see Table~\ref{tab:model_params}).  

\begin{figure}[!h]
\centering
\includegraphics[width=\columnwidth]{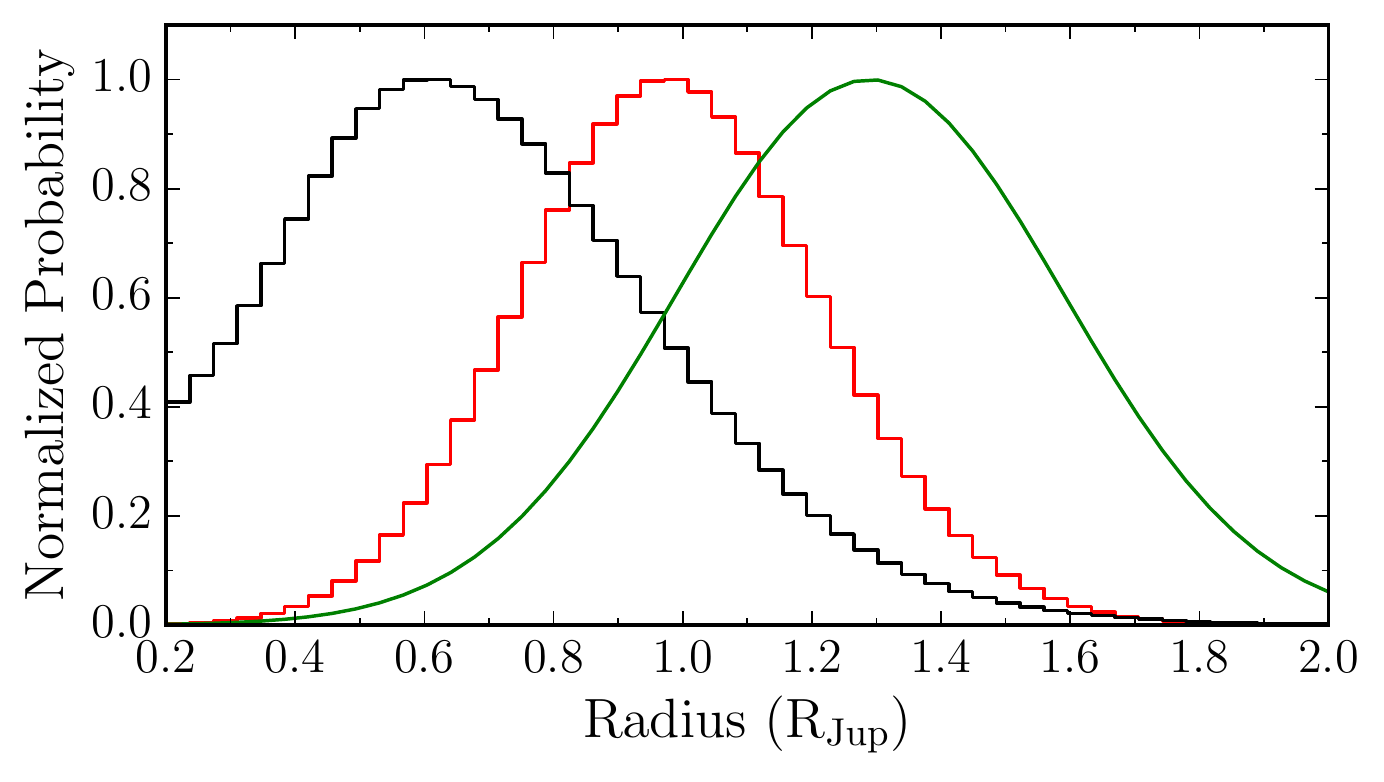}
\caption{ \label{fig:radpos} The figure shows the effect of applying a Gaussian radius prior when modeling with the iron/silicates grid. The prior shown by the green line is centered on the radius given by evolutionary models i.e. 1.29 R$_\text{Jup}$ \citep{Marley:2007, Fortney:2008}. Also plotted are the likelihood (black) and posterior distribution (red). }
\end{figure}

Presented in Figure~\ref{fig:MM_modelfits} is the best fitting model to the SED of 51~Eri~b. Stated in the figure are the model parameters along with the radius of the planet required to scale the model spectrum to match the planet SED. This scaling factor is required since the model spectra are typically computed to be the emission at the photosphere or at 10pc from the object. One of the free parameters in most model fitting codes is the term R$^2$/d$^2$ to scale the model flux to match the SED, where R is the radius of the planet and d is the distance to the object. For 51~Eri, the distance is known to better than 2\% (see Table~\ref{tab:props}) and thus we only fit the radius term. Shown in Figure~\ref{fig:MM_posterior} is the posterior distribution for the radius where we find that the best fitting radii are significantly smaller than that predicted by evolutionary models, e.g. 1.33--1.14 R$_\text{Jup}$ for a 2--10~M$_\text{Jup}$ hot/cold start planet at the age of 51~Eri \citep{Marley:2007, Fortney:2008}. This discrepancy has been noted previously as well for the HR8799 planets \citep{Marois:2008, Bowler:2010, Barman:2011, Currie:2011, Marley:2012}, $\beta$~Pic~b \citep{Morzinski:2015} and for 51~Eri~b itself in the discovery paper \citep{Macintosh:2015}. In an attempt to circumvent this issue, while modeling the SED we adopted a Bayesian prior probability density function for the radius in the form of a Gaussian centered on the expected radius from evolutionary models (green line in Figure~\ref{fig:radpos}), with the width chosen to include the radius of Jupiter. Without the prior (i.e. using a uniform prior), the median radius is 0.68 R$_\text{Jup}$ and T$_\text{eff} \sim 740$K, with the prior the median radius value is forced closer to the predictions of evolutionary models (red line in Figure~\ref{fig:radpos}) at 0.98 R$_\text{Jup}$, and T$_\text{eff}$ $\sim$ 690K, biasing the luminosity of the planet to larger values. When fitting the SED, the term that is conserved is the luminosity rather than the effective temperature or the radius. Adopting the evolutionary radius and marginalizing over the uncertainty in radius raises the luminosity ($\log L/L_\odot$) from -5.83 to -5.65. Since observational constraints on the radius for young planets are unavailable, we chose to use an uninformative prior.

\begin{figure*}[!ht]
\centering
\includegraphics[width=\linewidth]{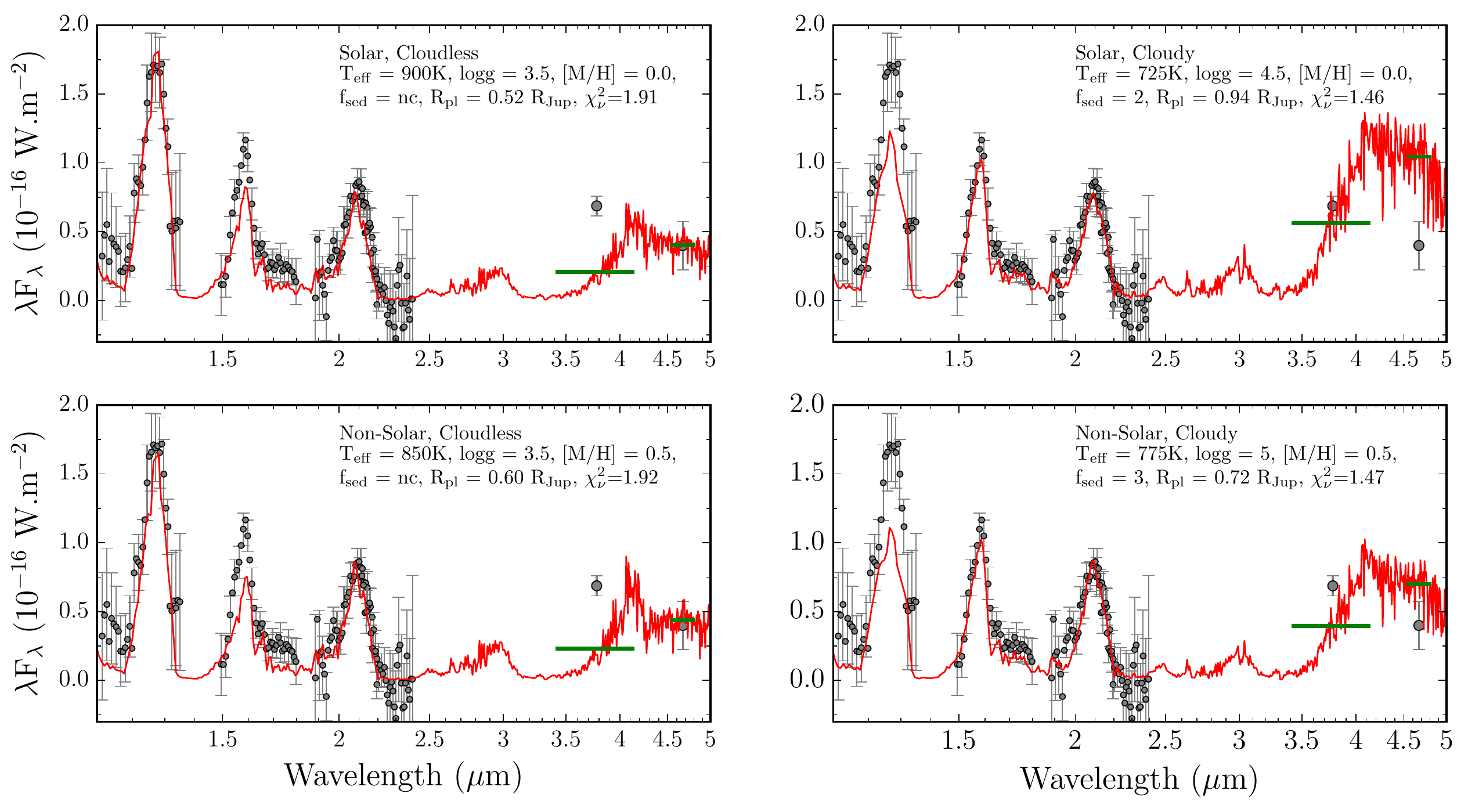}
\caption{ \label{fig:CM_modelfits} Spectral Energy Distribution of 51~Eri~b with the best fitting salt and sulfide cloud models. Each panel shows the best fitting model under the specific conditions: top two panels show the best fitting \emph{solar} metallicity models with cloudless atmosphere on the left and cloudy atmospheres on the right. Bottom two panels show the best fitting \emph{non-solar} models with cloudless atmosphere on the left and cloudy atmospheres on the right. }
\end{figure*}

Plotted in Figure~\ref{fig:MM_posterior} are the normalized posterior distributions for each of the model parameters varied in the model fit, along with the covariances to show how each of the parameters are affected. Since the grid only had a few models with $\log g$ = 3.5, with the majority being 3.25, we marginalized over the the surface gravity. The irregular shape of the effective temperature posterior is caused by the missing models in the grid. The median effective temperature, 737~K, estimated from the grid falls right in between the range of best fitting temperatures from the models in the \citet{Macintosh:2015} paper (700--750K). However, based on the shape of the posterior and the covariances, the peak of the effective temperature distribution extends to cooler temperatures. Since the L to T transition has been suggested to arise from holes or low opacity patches appearing in an initially more uniform cloud deck \citep{Ackerman:2001, Burgasser:2002, Marley:2010}, our finding here that partly cloudy models best fit the 51 Eri b spectrum is consistent with this interpretation. In general, however the models struggled to fit the entire planet SED, typically being able to fit either the near or mid IR portions of the SED. The inability to fit mid-IR photometry suggests that chemical equilibrium models are not appropriate. Disequilibrium chemistry predicts less CH$_4$ in the atmosphere and could explain higher flux at 1.6$\mu$m and in the $L_P$ band. It would also introduce CO, accounting for lower flux in the $M_S$ band.

\subsection{Sulfide and Salt Cloud Models}

In Section~\ref{ssec:comp}, we showed that the best fitting spectral type of 51~Eri~b is a mid-to-late T-dwarf. At the effective temperatures of mid to late T-dwarfs, Cr, MnS, Na$_2$S, ZnS, and KCl are expected to condense and form clouds high in the photosphere. The second grid we tested the planet SED against made use of a model grid which includes salt and sulfide clouds to test additional parameters such as the surface gravity and metallicity (which were varied, unlike the iron/silicates grid) and the properties of clouds typically associated with T-dwarfs. The grid was designed specifically for lower temperature objects \citep[$450 \sim 900$ K][]{Morley:2012, Morley:2014} and has been successfully to reproduce the SED of GJ~504~b \citep{Skemer:2016}, a cool low mass companion with a similar spectral type (late-T) which is comparable to 51~Eri~b \citep{Kuzuhara:2013}. Note that the use of this cloud grid does not preclude the possibility of the planet transitioning from L-to-T.

Also included as part of this grid are the clear atmosphere models from \citet{Saumon:2008}, the ranges for which are presented in Table~\ref{tab:model_params}. The range of parameters varied are presented in Table~\ref{tab:model_params}, including temperatures, surface gravities, metallicities, and sedimentation factor (f$_\text{sed}$) ranging from cloudy (f$_\text{sed}$ = 1) to cloud free. The cloud model used in the sulfide/salt grid is the same as the one described above. In addition to the opacity updates mentioned above, opacity effects due to alkali metals (Li, Na, K) have been included using the results from \citet{Allard:2005}. Between effective temperatures of 450--775 K, the grid is complete with models available for every step of the varied parameters. For effective temperatures between 800--900K, the temperature steps switch from increments of 25K to 50K and there are no models with f$_\text{sed}$ values of 1 and 2. This grid does not include opacity effect due to iron and silicates condensates. A future series of paper describing an extended atmosphere model grid will describe the updates, however the present grid extends the models to greater than solar metallicites. 

\begin{figure*}[!ht]
\centering
\includegraphics[width=\linewidth]{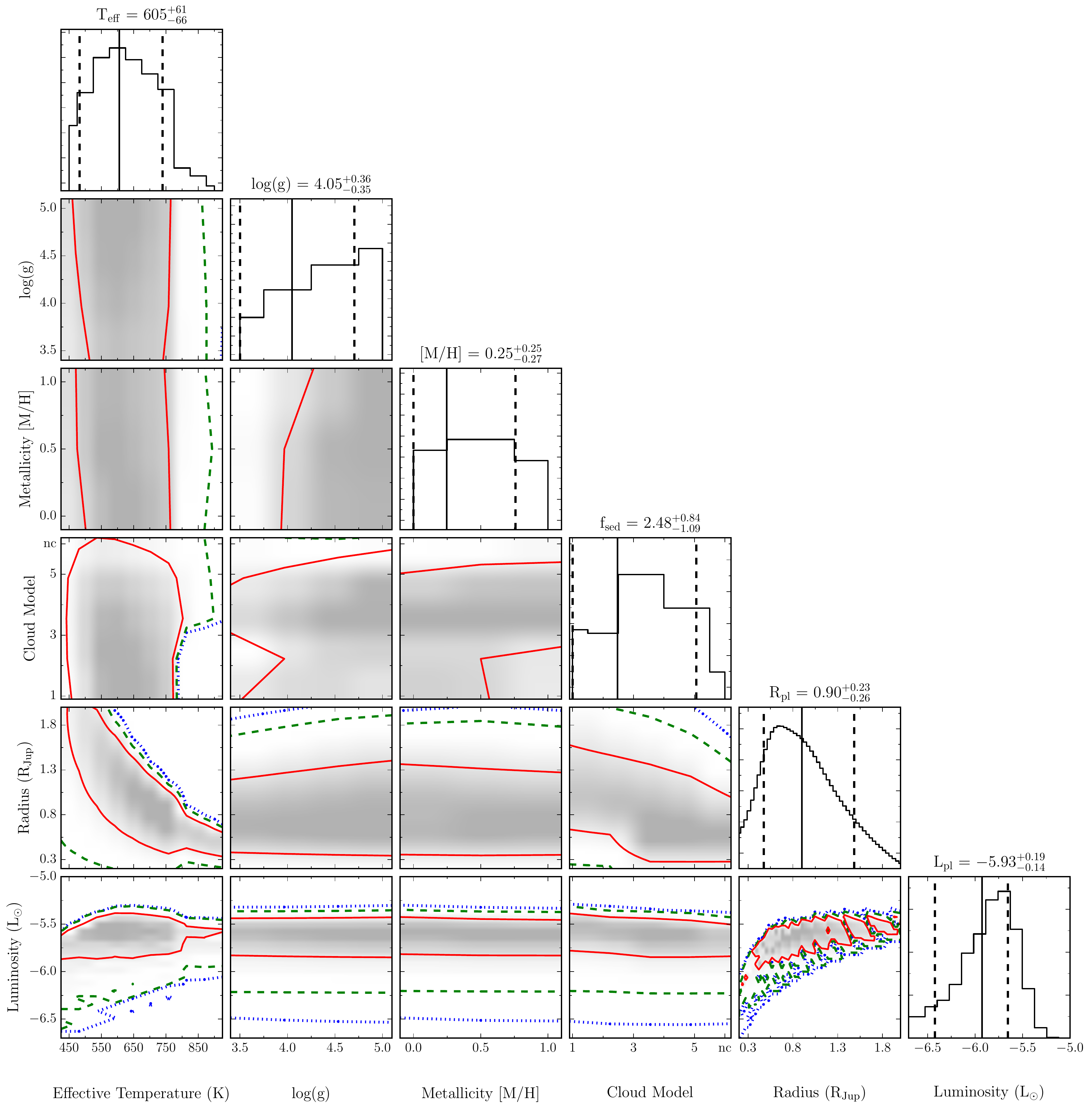}
\caption{ \label{fig:CM_posterior} Normalized posterior distributions for the sulfide and salt model grid. Same as Figure~\ref{fig:MM_posterior}.}
\end{figure*}

In Figure~\ref{fig:CM_modelfits}, we present the four best fitting model atmospheres for 51~Eri~b. Presented in each panel are the atmosphere with the lowest reduced chi-square in one of four cases, namely, solar and cloudless (top-left), solar and cloudy (top-right), non-solar and cloudless (bottom-left), non-solar and cloudy (bottom-right). Both cloudless model atmospheres are warmer and thus fit the near-IR spectrum of the planet while completely missing the $L_P$ photometry. The cloudy atmosphere model fits are cooler and do a much better job of fitting the overall SED of the 51~Eri~b and the best fitting atmosphere for both solar and non-solar metallicity have very similar reduced chi-square values. 

The normalized posterior distributions for the different parameters varied as part of the model fitting are shown in Figure~\ref{fig:CM_posterior}. The best fitting T$_\text{eff}$ (${605}_{-66}^{+61}$ K) is much cooler in comparison to the iron/silicates grid, but the values are within 2-$\sigma$ of each other. We also note out that the median might not be the best estimate for the effective temperature PDF in the iron/silicates grid where the peak extended to cooler temperatures. For the surface gravity and metallicity posterior distributions, we present the median values and error bar assuming a Gaussian distribution, though they may not be Gaussian. The surface gravity PDF suggests that the planet has high surface gravity. However 51~Eri~b is clearly a low mass companion indicating that the data does not constrain the gravity. A prior might help constrain the distribution, but there are currently no physically motivated priors available for the surface gravity of young planets. Similarly, the PDF for the metallicity is also unconstrained and higher resolution spectra in the $K$-band might help provide greater constraints on the metallicity of 51~Eri~b \citep{Konopacky:2013}. 

A difference between the iron/silicate and salt/sulfide atmosphere grids is in the planet radius, where the best fit radii for the cloudy models and the median radius of the PDF for the salt/sulfide models are much closer to evolutionary model predictions. A possible explanation for this discrepancy is that fitting the lower effective temperatures while still matching the bolometric luminosity, requires a larger radius. If the iron/silicates models extended to lower temperatures, assuming the continued presence of these clouds at these colder temperatures, it is likely that the radius discrepancy would not be as apparent. The sedimentation factor was fixed (at f$_\text{sed}$=2) in the iron/silicates grids, but had varying hole fractions (h$_\text{frac}$). In the sulfide/salt grid, f$_\text{sed}$ was varied and the median value for the distribution is f$_\text{sed}$=2.48. If we equate the h$_\text{frac}$ from the iron/silicates model with the f$_\text{sed}$ as the physics controlling the emission of flux from the photosphere then for both model grids the best fitting models tend to be favoring the presence of clouds over cloud free atmospheres. Furthermore, in both cases the best fitting models were not the fully cloudy atmospheres, with the smallest h$_\text{frac}$/f$_\text{sed}$. While the cloud compositions in both models are different, fitting either grid require cloud opacity. This can be achieved in one of two ways: either make the deep iron/silicates clouds be very vertically extended (small f$_\text{sed}$) or introduce a new cloud layer in the form of the sulfide/salt clouds.

\begin{figure}[!h]
\centering
\includegraphics[width=\columnwidth]{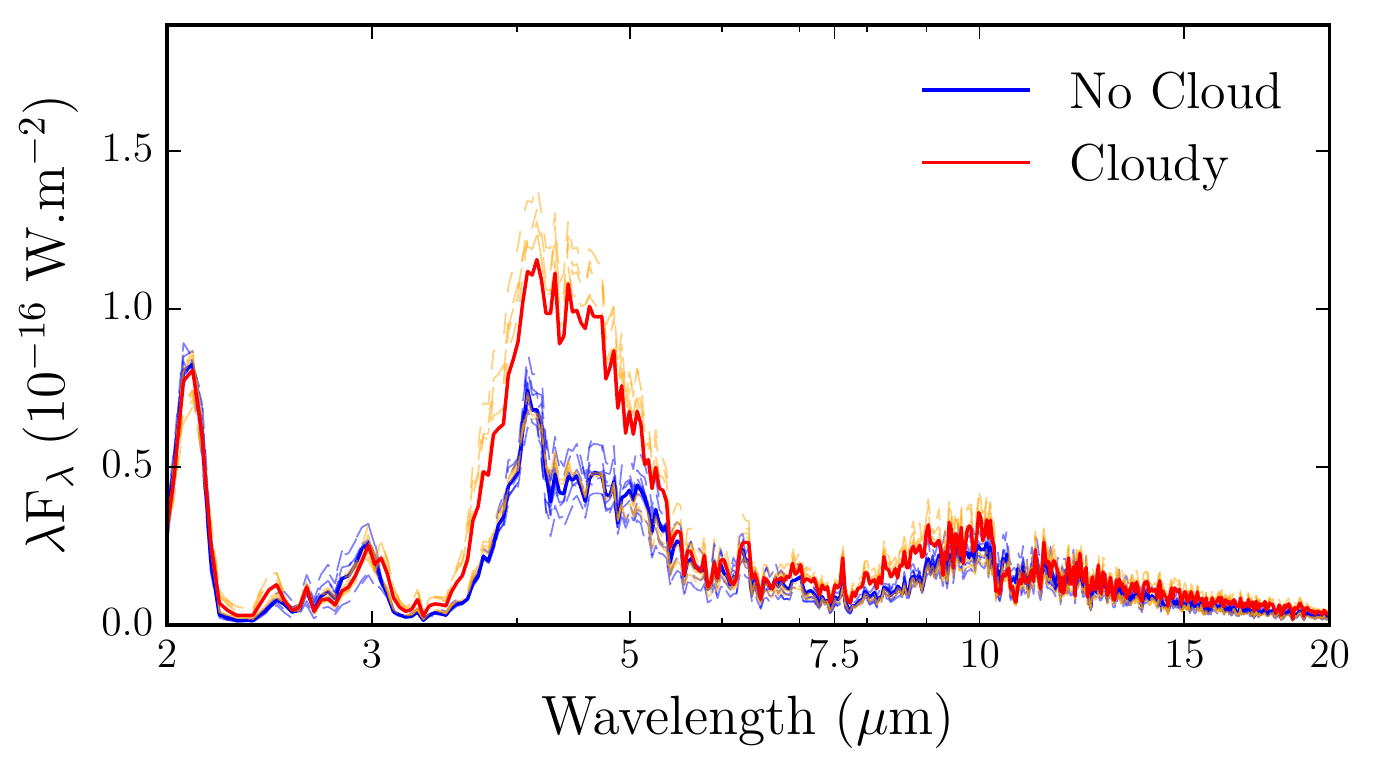}
\caption{  \label{fig:JWST} The ten best fitting cloudy (red) and cloudless (blue) atmospheres over the wavelength range of the \emph{James Webb Space Telescope}. The median of the models is plotted with a thicker line. The models indicate the divergence between the model fits over the wavelength covered by \emph{JWST}. }
\end{figure}

The cloudy model atmosphere fits presented in Figure~\ref{fig:CM_modelfits} match the $H$ through $K$ spectrum while being slightly under luminous in the $J$ and over luminous $M_S$ bands. Given the large photometric errors in the $M_S$ data, the model photometry lies within 2-$\sigma$ of the data. $JWST$ and other future low background mid-IR instruments will better constrain the 3--24~$\mu$m SED, a further test of current models. In Figure~\ref{fig:JWST}, we show ten of the best fitting models assuming cloudy (sulfide/salt clouds) or cloudless atmospheres extended out to 20$\mu$m. It is clear from these models that observations with the coronagraph on Near InfraRed Camera (NIRCam), spanning the 3--5~$\mu$m wavelength will add significant constraints on the atmosphere of the planet. If the planet can be studied with the Mid Infrared Instrument (MIRI), it could be used to apply constraints on chemical disequilibrium in the atmosphere through observations NH$_3$ in the 10--11$\mu$m range. 

\subsection{Luminosity of the planet}

The two different grids used in this study have produced similar luminosity predictions for the planet despite the different cloud compositions. From the iron/silicates grid we infer a bolometric luminosity of $\log L/L_\odot$ = $-5.83^{+0.15}_{-0.12}$, and $\log L/L_\odot$ = $-5.93^{+0.19}_{-0.14}$ from the sulfide/salt model atmospheres. We compare these luminosity estimates to predictions of evolutionary models to infer the planet mass and discuss its initial formation conditions.

\subsubsection{Standard cold- and hot-start models}

\begin{figure}[!h]
\centering
\includegraphics[width=\columnwidth]{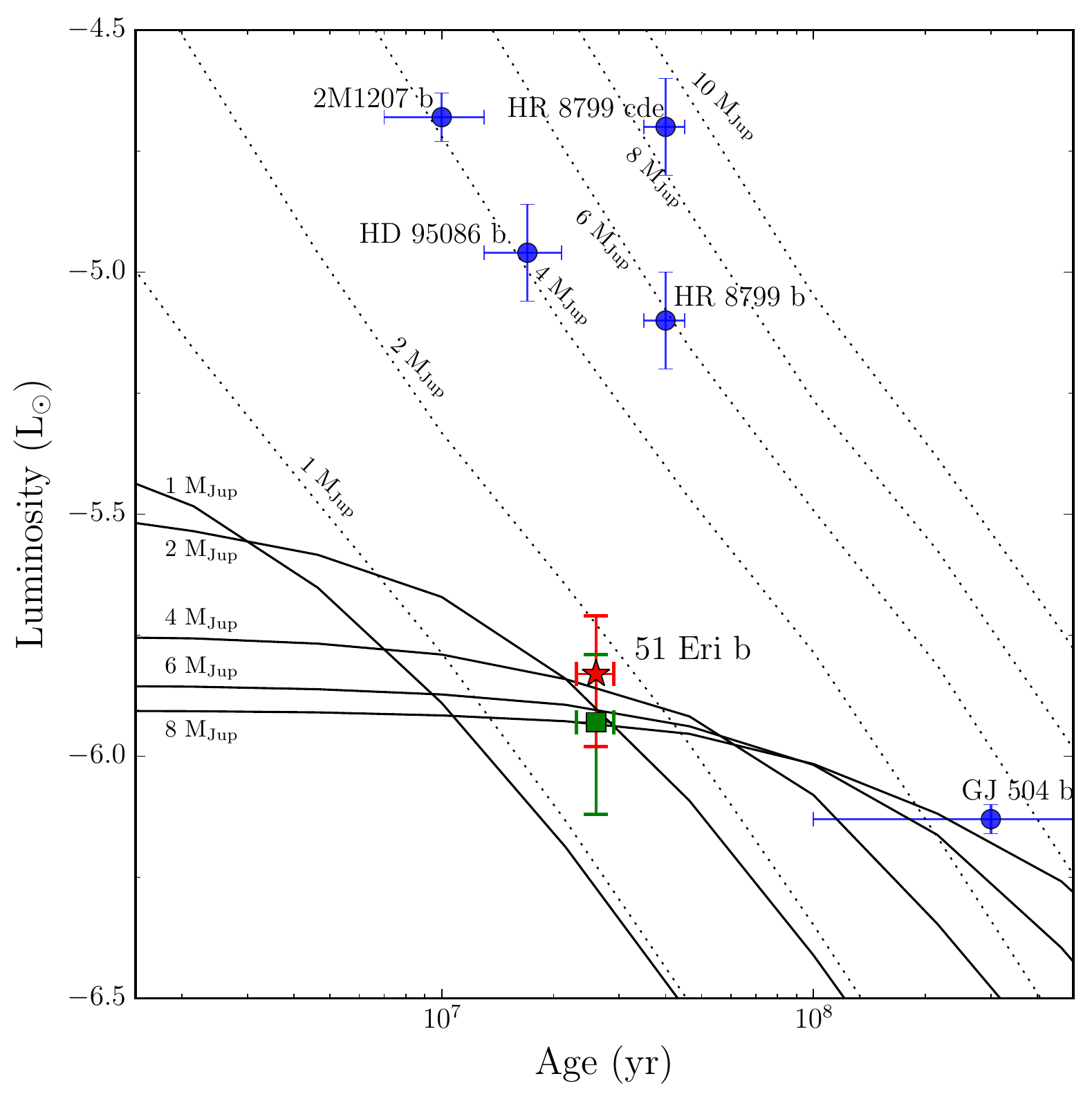}
\caption{  \label{fig:Lbol} Luminosity of imaged planetary mass companions as a function of age. For 51~Eri~b, the red star is the inferred luminosity from the Iron/Silicates model grid while the green square is the inferred luminosity from the Sulfide/Salt model grid. Also plotted are evolutionary tracks assuming different starting conditions i.e. the ``hot-start'' (dotted lines) and ``cold-start'' (solid lines) models of \citet{Marley:2007, Fortney:2008}. 51~Eri~b is consistent with both hot- and cold-start formation models. A subset of known directly imaged companions are plotted in figure illustrating the difference between 51~Eri~b and other imaged planets. Data for the companions, 2M1207~b, HR8799~bcde, HD~95086~b, GJ~504~b were taken from \citet{Patience:2010,Rajan:2015,Zurlo:2016,DeRosa:2016,Skemer:2016} respectively.}
\end{figure}

\begin{figure*}[!ht]
\centering
\includegraphics[width=\linewidth]{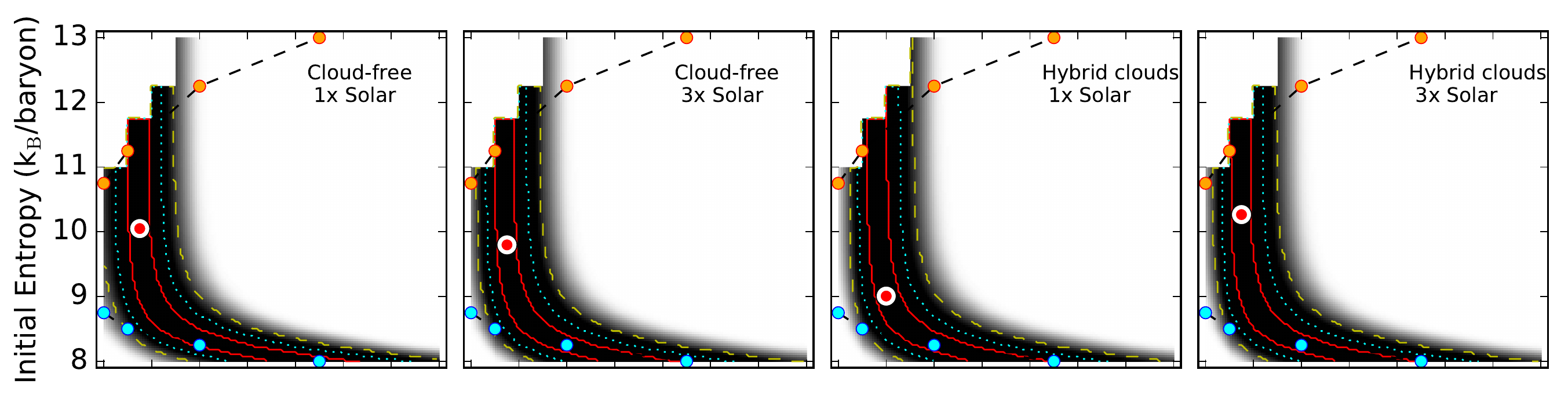}
\includegraphics[width=\linewidth]{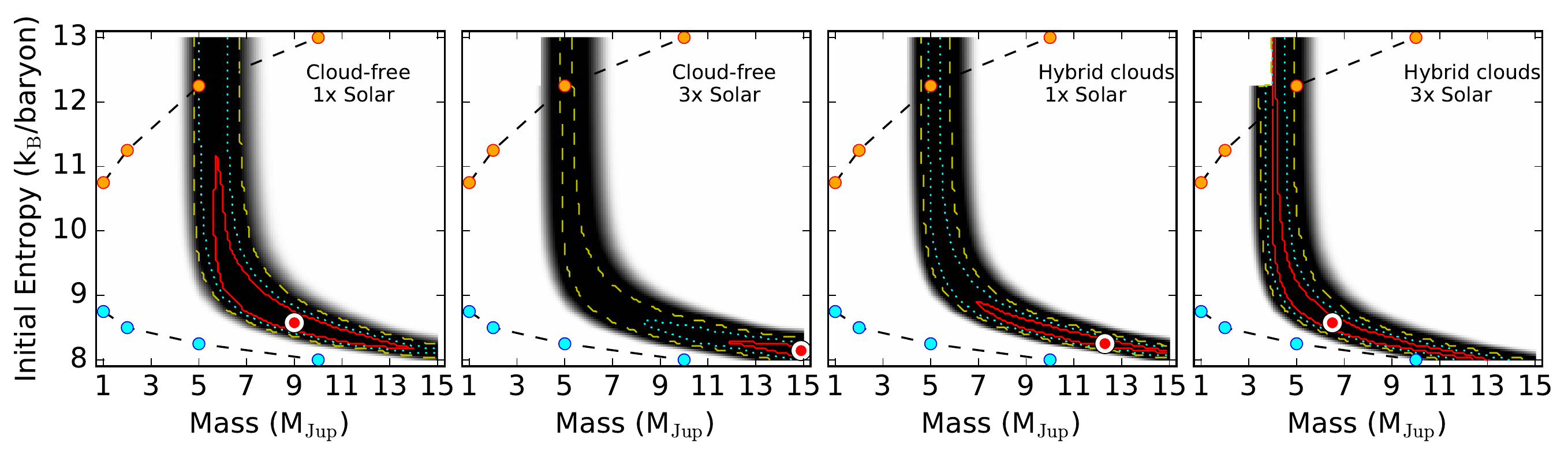}
\caption{  \label{fig:SvM} Comparing the planet spectrum and luminosity to a combination of initial entropy (k$_\text{B}$/baryon) and planet mass (M$_\text{Jup}$) from the warm-start evolutionary models of \citet{Spiegel:2012}. The four different atmospheres tested include cloud-free and hybrid cloud models, with both solar and super-solar metallicity. Also plotted are the 1$\sigma$ (solid red), 2$\sigma$ (dotted cyan) and 3$\sigma$ (dashed yellow) contours. The entropy plotted in the figure and used in the modeling, is not the entropy for the evolved object but rather the entropy at formation. The best fitting model fit is indicated by the large circle (white and red circle). The orange filled circles show the hot-start model limits, while the blue filled circles show cold-start, which are presented as the boundary cases in \citet{Spiegel:2012}. The top row is comparing the model luminosity to the inferred luminosity for 51~Eri~b, and the bottom row compares directly the SED to the evolutionary model spectra. }
\end{figure*}

In Figure~\ref{fig:Lbol} we compare the bolometric luminosity to evolutionary models for planets formed via the two extreme scenarios namely, hot-start and cold-start models \citep{Burrows:1997, Marley:2007}. In the hot-start scenario, planets are formed with high initial-entropy and are very luminous at birth. This scenario is usually associated with rapid formation in the circumstellar disk through disk instabilities. Alternatively, in the cold-start scenario, which is often associated with current 1D models of the core-accretion mechanism, planets start with a solid core that accretes gas from the stellar disk. The accreting gas loses energy via a radiatively efficient accretion shock and form with low initial-entropy and thereby lower post-formation luminosity. 

The other directly imaged companions plotted in Figure~\ref{fig:Lbol} can all be considered as having formed via the hot-start scenario. Despite the older age assessment for the companion in this study 26$\pm$3~Myr \citep{Nielsen:2016} compared to 20$\pm$6~Myr \citep{Macintosh:2015}, the revised luminosity when compared to the system age places 51 Eri b in a location where either cold or hot initial conditions are possible. Based on the hot-start tracks, it would have an inferred mass between 1--2 M$_\text{Jup}$. However, for the cold-start case the planet mass could lie anywhere between 2--12 M$_\text{Jup}$, since the model luminosity is largely independent of mass at the age of 51~Eri~b. Dynamical mass estimates for the planet could help clarify the formation mechanism especially if the planet mass $>2$M$_\text{Jup}$.

\subsubsection{Warm-start models}

\citet{Spiegel:2012} proposed a complete family of solutions existing between the hot- and cold-start extreme cases. Warm-start models\footnote{\url{http://www.astro.princeton.edu/~burrows/}} explore a wide range of initial entropies aimed at covering the possible range of initial parameters that govern the formation of planets. In Figure~\ref{fig:SvM}, we compare the inferred bolometric luminosity and the planet SED to models from \citet{Spiegel:2012}. The \citet{Spiegel:2012} models are evolutionary tracks calculated assuming different initial entropies for the planet, between 8 and 13 k$_\text{B}$/baryon, where k$_\text{B}$ is Boltzmann's constant, with steps of 0.25 k$_\text{B}$/baryon and masses between 1 and 15 M$_\text{Jup}$ with steps of 1 M$_\text{Jup}$. Four different model atmospheres are considered in combination with the evolutionary model: cloud-free and solar metallicity to fully cloudy with 3$\times$ solar metallicity \citep{Burrows:2011}. The bolometric luminosity of each point in the grid for each of the four atmosphere scenario was computed by integrating the SED over the wavelength range. Because of the sparse sampling of the grid, we linearly interpolate the evolutionary tracks with steps of 0.06 k$_\text{B}$/baryon and 0.2 M$_\text{Jup}$. 

In the top row of Figure~\ref{fig:SvM}, we plot the probabilities for each grid point measured by comparing the average of the inferred bolometric luminosities from the SED fit ($\log L/L_\odot$ = $-5.87\pm0.15$) to the predictions of the \citet{Spiegel:2012} models with the four atmosphere conditions. For the bottom row in Figure~\ref{fig:SvM}, the surface is calculated by fitting the planet SED to the \citet{Spiegel:2012} model atmosphere grid, using Equation~\ref{chi-square}. For both comparisons, luminosity and SED, we chose the age of the evolutionary grid best matching the age of 51 Eri (25 Myr), to minimize the number of interpolations, and only varied the mass of the planet and initial entropy for the models.

\citet{Mordasini:2013} find that the luminosity of a planet that underwent accretion through a super-critical shock (the standard cold-start core accretion hypothesis), is highly dependent on the mass of the core, M$_\text{core}^{2-3}$. Therefore, the continuum of warm-start models can also be explained by similar bulk mass planets with increasing core mass. These models suggest that the entropy of 51~Eri~b can be explained via core-accretion, with a core mass ranging between 15 and 127~M$_{\oplus}$, which can reproduce the planet luminosity with various initial entropies.

The four panels generated by fitting the inferred luminosity (upper four panels) appear highly consistent and in agreement with the results from Figure~\ref{fig:Lbol}. The 1$\sigma$ contour encompasses the entire available entropy space, where for intermediate and high entropies the most likely mass for the planet is between 2 and 3 M$_\text{Jup}$ and for low initial entropy the most likely mass for the planet increases, making distinguishing between cold-, warm- and hot-start difficult. 

When we compare the model spectra directly to the planet SED, the surface is qualitatively similar to that made with the luminosity but shifted to higher mass and with the 1$\sigma$ contours and best fit models favoring lower entropy. According to the \citet{Mordasini:2013} models, the fits presented here would be consistent with a planet having core masses ranging from 15--127~M$_{\oplus}$. 

Conversely to other directly imaged companions \citep[see figures in][]{Marleau:2014}, 51~Eri~b is the only planet compatible with very low initial entropy and the cold-start case. Tighter constraints on the bolometric luminosity and/or higher signal to noise data will help to reduce the width of the two branches and independent mass constraints, from dynamical measurements, will enable to infer the initial entropy and possible formation route. Atmospheric retrievals and/or higher resolution spectra aimed at exploring and characterizing the planets chemical composition might also help understand whether the planet has higher C/O ratios compared to the star, since planetary C/O can be used to understand planet formation \citep{Oberg:2011, Konopacky:2013}.

\section{Conclusion}

In this paper, we have presented the first spectrum of 51~Eridani~b in the K-band obtained with the Gemini Planet Imager (K1 and K2 bands) as well as the first photometric measurement of the planet at $M_S$ obtained with the NIRC2 Narrow camera. We also obtained an additional $L_P$ photometric point that agrees very well with the $L_P$ measurement taken in the discovery paper \citep{Macintosh:2015}. In addition, we revised the stellar photometry by observing the star in the near IR and estimating its photometry in the mid IR through an SED fit. The new data are combined with the published $J$, and $H$ spectra and the $L_P$ photometry to present the spectral energy distribution spanning 1--5~$\mu$m for the planet.

As part of the data analysis, we calculated the covariance for each of the spectral datasets i.e. $J, H, K1$, and $K2$ using the formalism presented in \citet{Greco:2016}. The spectral covariance was used in all the chi-squared minimization performed as part of this study, in combination with the photometric variance. Using the covariance ensured that the photometric points were weighted in a suitable manner and resulted in cooler effective temperatures for the best fits.

We compared the planet photometry to field and young brown dwarfs by fitting their near-IR spectra to 51~Eri~b to estimate a spectral type of T$6.5\pm1.5$. Due the relative paucity of known young T-dwarfs, our comparison of the planet spectrum to young T-dwarfs only included a handful of objects, and amongst the sample 51~Eri~b appears to have the lowest surface gravity based on a comparison of their spectral shape and amplitude.

In a comparison of the near and mid IR photometry for the planet to the field and young brown dwarf population via a range of color magnitude diagrams we note that 51~Eri~b is redder than brown dwarfs of similar spectral types. This was also noted in the discovery paper, and it was proposed that this might be due to presence of clouds, similar to young L-type planetary mass companions. In this study, we extended this idea to suggest that a possible reason for the presence of clouds (compared to the field), is that the planet is still transitioning from the L-type to the T-type. This would occur at a lower $J$ magnitude than field brown dwarfs due to its lower mass when including a gravity-dependent transition in the evolution \citep{Saumon:2008}. 

We also fit the planet SED with two different model atmosphere grids that varied in the composition of molecules that could condense in the atmosphere. The best fitting models in both cases, were those that contained large amount of condensates in the atmosphere as compared to cloud free atmospheres. Through the iron/silicates grid, we estimate that the planet has a patchy atmosphere with 10--25 \% hole fraction in the surface cloud cover, which is consistent with the f$_\text{sed}$ values of 2--3 resulting from the sulfide/salt grid. The median effective temperature from the two grids is ${737}_{-46}^{+39}$~K and ${605}_{-66}^{+61}$~K for iron/silicates and sulfide/salt respectively. This value is slightly cooler, compared to \citet{Macintosh:2015}, where the best fit models had temperatures of 700K and 750K respectively. The surface gravity and metallicity both appear to be unconstrained by the data, but empirical fits to young T-dwarfs suggest that the planet has lower surface gravity. 

The two atmosphere grids provide similar luminosity estimates which were compared to hot-, warm- and cold-start models. 51~Eri~b appears to be one of the only directly imaged planet that is consistent with the cold-start scenario and a comparison of the planet SED to a range of initial entropy models indicates that cloudy atmospheres with low initial entropies provide the best fit to the planet SED. 

Following the submission of this study for publication, a paper on 51~Eri~b using spectrophotometry taken with the VLT/SPHERE was published by \citet{Samland:2017}. Their study includes new $YJH$ spectra as well as $K1K2$ photometry in addition to the $H$ spectrum and $L_P$ photometry from \citet{Macintosh:2015}. Their results are consistent in parts with ours, although we note that the SPHERE $J$ band spectrum is fainter than the GPI $J$ spectrum, while their $K1K2$ photometry are brighter than the GPI spectrum (and corresponding integrated GPI photometry). These differences could very well be caused by the application of different algorithms, where \citet{Samland:2017} demonstrate that different algorithms can result in spectra with a range of flux values including ones that agree with the GPI $J$ spectrum. Future studies will need to analyze all the available datasets using a common pipeline for data processing and analysis to understand whether the differences arise from the algorithms or due to other causes. 

With future space missions such as the \emph{James Webb Space Telescope}, the 3--24~$\mu$m~SED of this planet could be observed at higher SNR, providing tests of current atmospheric models. The best fitting atmosphere models further indicate that the planet might have a cloudy atmosphere with patchy clouds, making 51~Eri~b a prime candidate for atmospheric variability studies that might be possible with future instrumentation. Further analysis of this data using methods such as atmosphere retrievals could permit an exploration of other planet parameters that were not considered in this study such as chemical composition of the atmosphere and the thermal structure.

\acknowledgments
{\bf Acknowledgements:} The authors thank Gabriel Marleau for discussion on warm-start models. The Gemini Observatory is operated by the AURA under a cooperative agreement with the NSF on behalf of the Gemini partnership: the National Science Foundation (United States), the National Research Council (Canada), CONICYT (Chile), the Australian Research Council (Australia), Minist\'erio da Ci\'encia, Tecnologia e Inova\c{c}\=ao (Brazil), and Ministerio de Ciencia, Tecnolog\'ia e Innovaci\'on Productiva (Argentina). Supported by NSF grants AST-1411868 (A.R, J.L.P., B.M.), AST-1518332 (R.J.D.R.), and DGE-1311230 (K.W.D.). F.M and E.N are supported by NASA Grant NNX14AJ80G. Supported by Fonds de Recherche du Qu\'ebec (J.R, R.D, D.L). K.M.M and T.S.B are supported by the NASA Exoplanets Research Program (XRP) by cooperative agreement NNX16AD44G. G.V and J.K.W acknowledge JPL?s ESI program for GPI related funding. The results reported herein benefitted from collaborations and/or information exchange within NASA's Nexus for Exoplanet System Science (NExSS) research coordination network sponsored by NASA's Science Mission Directorate and the NExSS grant NNX15AD95G. Portions of this work were performed under the auspices of the U.S. Department of Energy by Lawrence Livermore National Laboratory under Contract DE-AC52-07NA27344. This research has benefited from the SpeX Prism Library (and/or SpeX Prism Library Analysis Toolkit) maintained by Adam Burgasser at \url{http://www.browndwarfs.org/spexprism}, the IRTF Spectral Library maintained by Michael Cushing, and the Montreal Brown Dwarf and Exoplanet Spectral Library maintained by Jonathan Gagn{\'e}.

\vspace{5mm}
\facilities{Gemini:South (GPI), Keck:II (NIRC2)}

\software{GPI Data Reduction Pipeline \citep[v1.3.0;][]{Perrin:2014}, pyKLIP \citep{Wang:2015}.}

\appendix

\section{Derivation of Spectral Covariance}
\label{app:cov2}
We follow the method described in \citet{Greco:2016} to measure the inter-pixel correlation within the PSF-subtracted images, and convert these into a covariance matrix. For each image ($J$, $H$, $K1$, and $K2$), the correlation $\psi_{ij}$ between pixel values at wavelengths $\lambda_i$ and $\lambda_j$ within a 1.5 $\lambda/D$ annulus was estimated as
\begin{equation}
    \psi_{ij} = \frac{\langle I_i I_j\rangle}{\sqrt{\langle I^2_i \rangle \langle I^2_j \rangle}}
\end{equation}
where $\langle I_i\rangle$ is the average intensity within the annulus at wavelength $\lambda_i$. This was repeated for all wavelength pairs, and at five different separations: 350, 454 (the separation of 51 Eri b), 550, 650, and 750~mas. To avoid biasing the measurement, 51~Eri~b was masked in the 454~mas annulus.

\begin{figure}
\centering
\includegraphics[width=12cm]{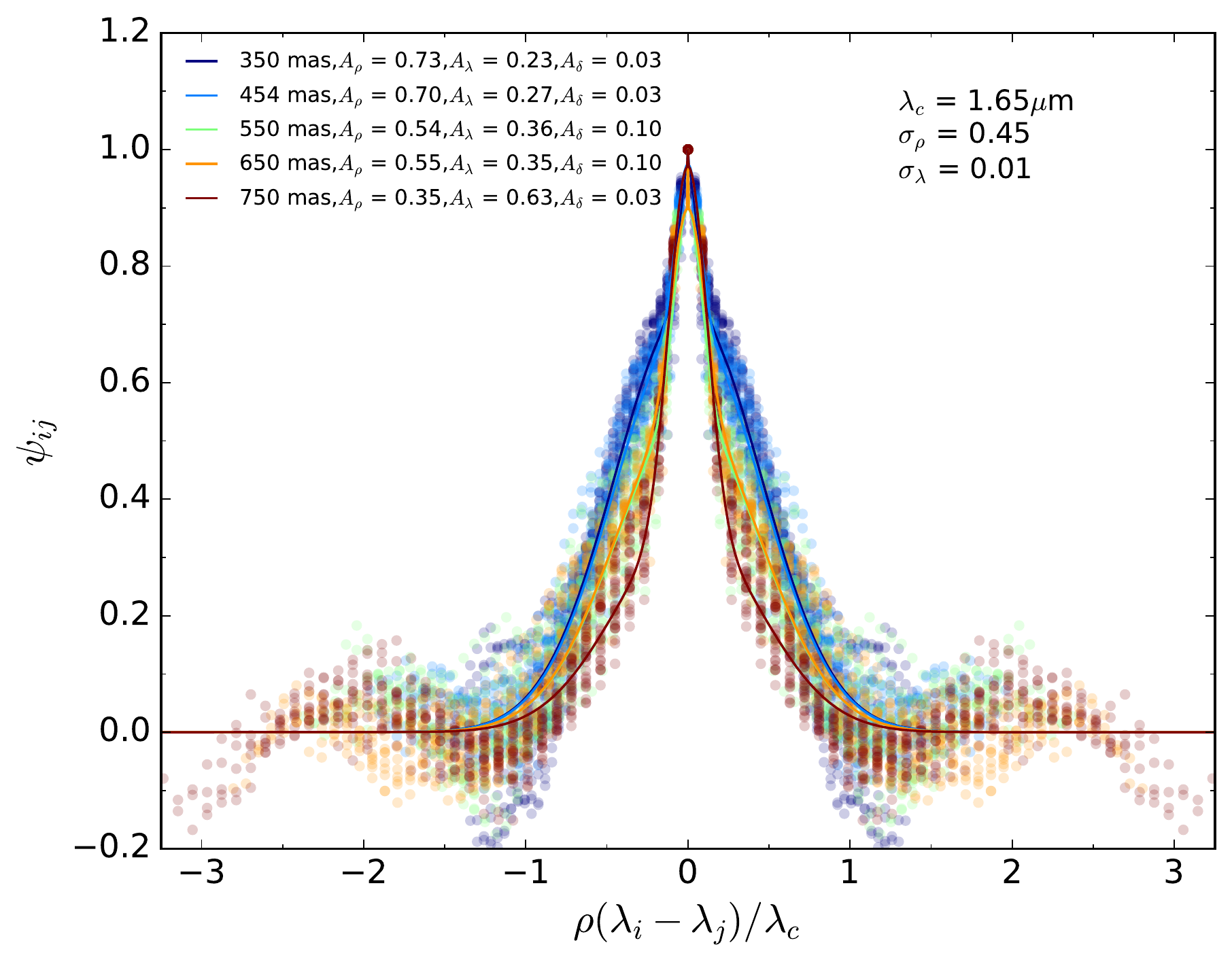}
\caption{ \label{fig:psisep} Example of the correlation function at the various angular separations included in the fit for the $H$-band spectral cube. The different colors correspond to the angular separations, with the circles being the value of the correlation for all the wavelength pairs and the lines of the same color indicate the best fit to distribution calculated using Equation~\ref{eqn:model}. }
\end{figure}

\begin{figure}
\centering
\includegraphics[width=10cm]{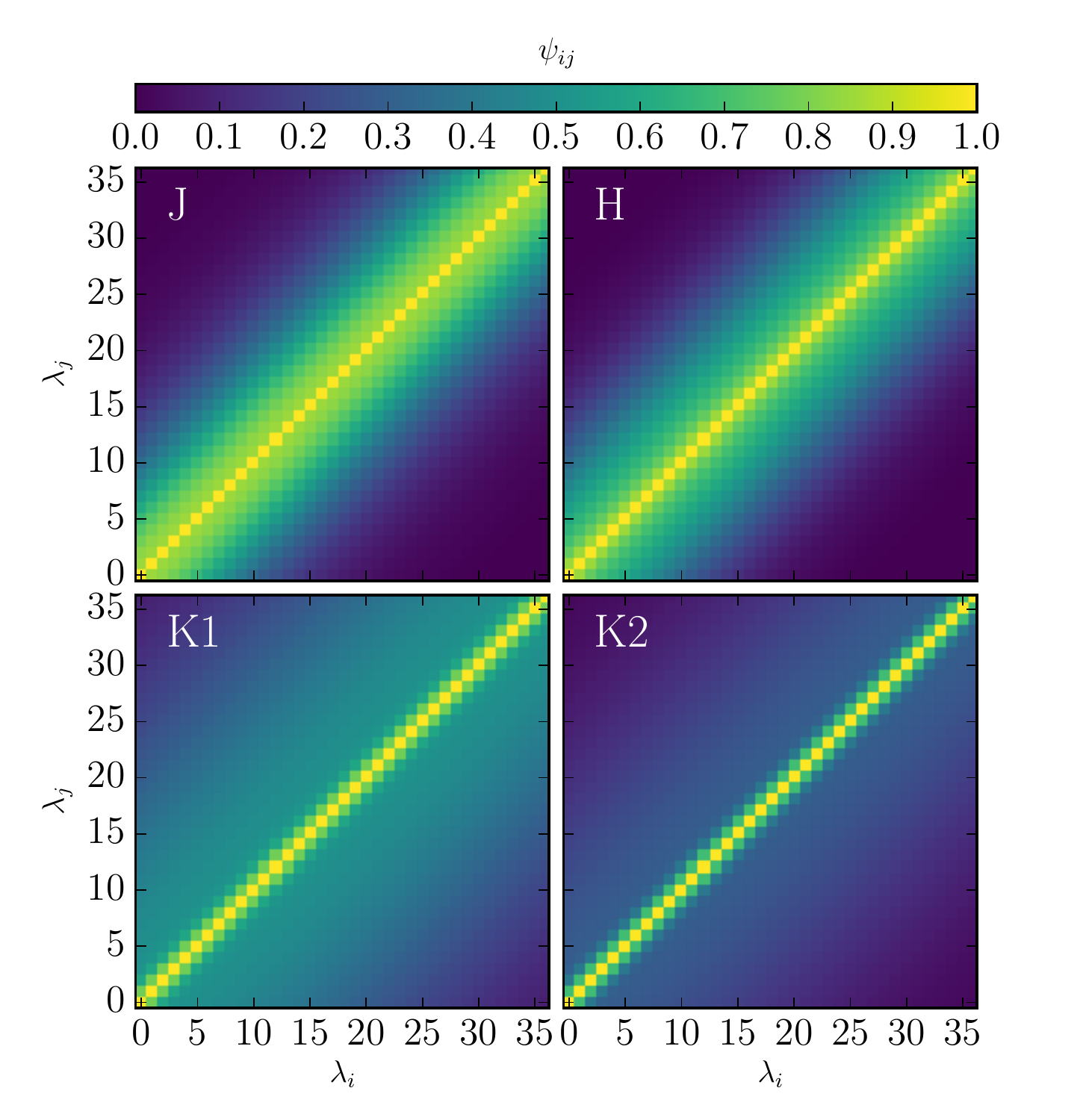}
\caption{ \label{fig:psi} Presenting the correlation matrices calculated for each of the four $JHK1K2$ spectra. Going from $J$-band through $K2$ the correlation length can be seen to change as a function of speckle vs background noise. The spectra are highly correlated at $J$ with up to 5 channels showing high correlation values, down to $\sim$3 at $K2$ which is a consequence of the spectral re-sampling.}
\end{figure}

The measurements of the correlation $\psi_{ij}$ at the eight different separations within the final image were used to fit the parametrized correlation model of \citet{Greco:2016},
\begin{equation}
\label{eqn:model}
    \psi_{ij}\approx A_{\rho}\exp\left[-\frac{1}{2}\left(\frac{\rho}{\sigma_{\rho}}\frac{\lambda_i - \lambda_j}{\lambda_c}\right)^2\right] + A_{\lambda}\exp\left[-\frac{1}{2}\left(\frac{1}{\sigma_{\lambda}}\frac{\lambda_i - \lambda_j}{\lambda_c}\right)^2\right] + A_{\delta}\delta_{ij}
\end{equation}
where the symbols are as in \citet{Greco:2016}. This model is based on the assumption that the correlation consists of three components. The first two terms model the contribution of the speckle noise and the correlation induced by the interpolation within the reduction process. The third models uncorrelated noise, such as read noise, which do not contribute to the off-diagonal terms of the correlation matrix. The amplitude of the first two terms ($A_{\rho}$, $A_{\lambda}$) were allowed to vary with separation, while the two correlation lengths ($\sigma_{\rho}$, $\sigma_{\lambda}$) were fixed. As the sum of the amplitudes must equal unity, $A_{\delta}$ was derived from the other amplitudes. Figure~\ref{fig:psisep} shows an example of the spectral correlation as a function of the angular separation for the $H$-band spectral cube, $\lambda_c$ is the central wavelength of the spectrum (1.65~$\mu$m for $H$). The colored lines in the plot are the best fits to Equation~\ref{eqn:model}. 

Due to the high dimensionality of the problem, we use a parallel-tempered Markov Chain Monte Carlo algorithm \citep{Foreman-Mackey:2013} to find the global minimum. The best fit parameters at the separation of 51~Eri~b within the PSF-subtracted image at each band is given in Table~\ref{tab:cov_params}. Using these parameters, the covariance matrix, $C$, was constructed for each band. The diagonal elements contained the square of the uncertainties of the spectrum of the planet, and the off-diagonal elements were calculated using
\begin{equation}
    \psi_{ij} \equiv \frac{C_{ij}}{\sqrt{C_{ii}C_{jj}}}
\end{equation}

\begin{deluxetable}{cccccccc}
\tablecolumns{6} 
\tablewidth{0pt} 
\tablecaption{Correlation model parameters \label{tab:cov_params}} 
\tablehead{ 
\colhead{Band}  & \colhead{$A_{\rho}$} & \colhead{$A_{\lambda}$} & \colhead{$A_{\delta}$} & \colhead{$\sigma_{\rho}$} & \colhead{$\sigma_{\lambda}$}}
\startdata 
$J$ & 0.43 & 0.43 & 0.14 & 0.44 & 0.05\\
$H$ & 0.70 & 0.27 & 0.03 & 0.45 & 0.01\\
$K1$ & 0.51 & 0.41 & 0.07 & 0.68 & 0.004\\
$K2$ & 0.30 & 0.62 & 0.08 & 0.43 & 0.004\\
\enddata 
\vspace{-0.8cm} 
\end{deluxetable}

The fitted parameters in Table~\ref{tab:cov_params} demonstrate that the primary cause of correlation at the shorter wavelengths is speckle noise, with the correlation induced by interpolation becoming more significant in the $K1$ and $K2$ images. In each case the amplitude of the speckle noise term (A$_{\rho}$) is significantly higher than seen for HD~95086~b \citep{DeRosa:2016}. This can be attributed to the fact that 51~Eri~A is approximately two magnitudes brighter at $K1$ (than HD~95086~A), leading to a significantly brighter speckle field. The typical correlation lengths in the PSF-subtracted image for each band are visualized in Figure~\ref{fig:psi}, with the data being highly correlated at $J$ band at wavelengths separated by up to five spectral channels.

\end{document}